\documentclass[sigconf,nonacm]{acmart}

\usepackage{hyperref}
\usepackage{cleveref}
\usepackage[inline]{enumitem}
\usepackage{subfig}
\usepackage{amsmath}
\usepackage[ruled,linesnumbered]{algorithm2e}
\usepackage{adjustbox}
\usepackage{multirow}

\setcopyright{acmlicensed}
\copyrightyear{2024}
\acmYear{2024}
\acmDOI{XXXXXXX.XXXXXXX}

\acmConference[ACSAC 24]{Annual Computer Security Applications Conference}{December 09--13, 2024}{Waikiki, HI}

\acmISBN{978-1-4503-XXXX-X/18/06}

\begin{document}

\title{Model-agnostic clean-label backdoor mitigation in cybersecurity environments}

\author{Giorgio Severi}
\affiliation{%
\institution{Northeastern University}
\country{}
}
\author{Simona Boboila}
\affiliation{%
\institution{Northeastern University}
\country{}
}
\author{John Holodnak}
\affiliation{%
\institution{MIT Lincoln Laboratory}
\country{}
}
\author{Kendra Kratkiewicz}
\affiliation{%
\institution{MIT Lincoln Laboratory}
\country{}
}
\author{Rauf Izmailov}
\affiliation{%
\institution{Peraton Labs}
\country{}
}
\author{Michael J. De Lucia}
\affiliation{%
\institution{DEVCOM Army Research Laboratory}
\country{}
}
\author{Alina Oprea}
\affiliation{%
\institution{Northeastern University}
\country{}
}

\date{}

\begin{abstract}
The training phase of machine learning models is a delicate step, especially in cybersecurity contexts.
Recent research has surfaced a series of insidious training-time attacks that inject backdoors in models designed for security classification tasks without altering the training labels.
With this work, we propose new techniques that leverage insights in cybersecurity threat models to effectively mitigate these clean-label poisoning attacks, while preserving the model utility.
By performing density-based clustering on a carefully chosen feature subspace, and progressively isolating the suspicious clusters through a novel iterative scoring procedure, our defensive mechanism can mitigate the attacks without requiring many of the common assumptions in the existing backdoor defense literature.
To show the generality of our proposed mitigation, we evaluate it on two clean-label model-agnostic attacks on two different classic cybersecurity data modalities: network flows classification and malware classification, using gradient boosting and neural network models.
\end{abstract}


\maketitle

\section{Introduction}

Machine learning (ML) models power an ever growing variety of software systems, including cybersecurity solutions intended to stop active adversaries~\cite{antonakakis11,MADE, mirskyKitsuneEnsembleAutoencoders2018a,nazca_ndss2014,tamersoy14,microsoftdefender,ibmqradar,fireeye2020Jun,xdr_solutions}.
The deployment of models in security-sensitive contexts increases the relevance of adversarial machine learning risks, both during inference, \emph{evasion attacks}, and during the training phase of the model, \emph{poisoning attacks}.
Recent trends in ML practices, especially concerning the growing size of datasets and increased reliance on data crowd-sourcing \cite{carlini2024poisoning}, and the widespread adoption of models as core components in cybersecurity products have increased the public awareness \cite{sivakumarAdversarialMachineLearningIndustry2020,altasVTPoison} of risks associated with training time adversarial interference.

Existing research in this area of adversarial ML has primarily focused on the computer vision domain. Seminal works such as \cite{biggio2012poisoning, guBadNetsEvaluatingBackdooring2019}, demonstrated different adversarial poisoning  procedures designed for various model architectures and targeted at achieving different adversarial objectives.
However, recent efforts by researchers have started exploring possible strategies an adversary can use
to compromise the integrity of models developed for cybersecurity tasks through training process manipulation~\cite{severiExplanationGuidedBackdoorPoisoning2021,Jigsaw,severiPoisoningNetworkFlow2023}.

We focus on mitigating backdoor attacks \cite{guBadNetsEvaluatingBackdooring2019} in cybersecurity settings. This type of attack aims to induce a victim model to memorize the association between a predetermined data pattern – also called a \emph{trigger} or backdoor – selected by the adversary, and a target class of the attacker's choice.
If successful, the same pattern can then be presented to the model during inference within any arbitrary data sample, triggering the target class output. 

We select backdoor attacks as our objective because we argue they pose a particularly relevant threat to cybersecurity applications.
These systems 
rely on large datasets of labeled samples, often gathered through network monitoring or crowdsourced feeds, providing opportunities for training data manipulation.
Moreover, backdoor attacks are inherently stealthy.
The adversary's objective is not to simply disrupt the performance of a victim model, which would lead to prompt discovery and potential remediation by the model owners.
Instead, the goal is to embed an arbitrary behavior – the ability to trigger a desired output – in the learned model without altering its normal behaviors on standard data points.
This makes it particularly difficult for a defender to realize that the victim model has been compromised.

In this work, we consider two particularly insidious backdoor attacks developed specifically to target ML models designed for cybersecurity tasks: one aimed at subverting static malware classifiers \cite{severiExplanationGuidedBackdoorPoisoning2021}, and one geared towards automated traffic analysis systems \cite{severiPoisoningNetworkFlow2023}.
Both attacks operate in a clean-label fashion, injecting the trigger pattern only in a small amount of training points corresponding to benign data, without requiring control over the training set labels.
In addition, these types of attacks are agnostic to the victim's model type, and are applicable to a variety of model architectures and data modalities used for  security classification tasks.

Our defensive approach leverages the information asymmetry between attacker and defender in these scenarios to isolate the poisoned points while maximizing the amount of \emph{clean} data retained for model training.
Our method revolves around an iterative scoring process on clustered data points in a selected feature subspace, and we propose different analysis procedures to remediate the effect of the poisoned clusters, able to reduce attack success by up to 90\% while preserving high utility, even at large poisoning rates up to 5\%.
In contrast with most existing backdoor mitigation approaches, we do not make specific assumptions on the victim model's architecture, and our approach is applicable to any classifier (not just neural networks).
Moreover, our defense removes another standard assumption often found in the literature: requiring the defender to have access to a set of clean data points sampled from the same distribution as the training data~\cite{liu2018fine-pruning,Heng2023SEAM,qiProactiveMLApproach2023}.

To summarize, we make the following contributions:
\begin{itemize}
    \item We propose a novel defense mechanism against clean-label backdoor attacks in cybersecurity domains. Our technique does not require access to clean trusted data or knowledge of the victim's model architecture, thus removing some strong assumptions from previous works.
    \item We demonstrate that our defense is generally applicable across various data modalities (network flow traffic and binary files), and multiple model types, such as neural networks, gradient  boosting trees, and LightGBM classifiers.
    \item We comprehensively evaluate our techniques against existing backdoor attacks from previous literature. We show that our methods are able to reliably identify and remove backdoor examples  while preserving a high model utility (F1 score) and low false positive rates.
    \item For reproducibility, we evaluate our defense strategies on publicly available datasets and open source attack implementations, and will release all the code used to run the experiments in this paper. 
\end{itemize}

\section{Related work}
In this section, we provide related work on backdoor attacks and mitigations, with a particular focus in the cybersecurity domain.

\subsection{Backdoor attacks} 

Adversarial examples \cite{DBLP:journals/corr/SzegedyZSBEGF13}, also known as evasion attacks, are the most well known typology of attacks, and likely the first line of attack against modern machine learning models.
In this setting, imperceptible perturbations are applied to test samples in order to change the model's output label during deployment.
Due to their immediate relevance in adversarial contexts, these attacks have been studied in the security domain for a relatively long time~\cite{Handley2001,Cao2017,Ayub2020,papadopoulosLaunchingAdversarialAttacks2021,FENCE}.
More recently, however, poisoning attacks against ML, which manipulate the training set (or other training parameters), have emerged as a top concern in industry~\cite{sivakumarAdversarialMachineLearningIndustry2020}.
For a deep dive into adversarial machine learning techniques we direct the interested reader to the comprehensive standardized NIST taxonomy~\cite{opreaAdversarialMachineLearning2023}.

In this work we focus on \emph{backdoor} poisoning attacks, a notably dangerous technique that does not affect the model's predictions on clean test data; instead, by injecting a trigger pattern into a small fraction of the training data, the adversary is able to control the model's predictions whenever the same trigger appears during model deployment. In their seminal work, Gu et al.~\cite{guBadNetsEvaluatingBackdooring2019} used a small pixel-based trigger pattern to misclassify images of handwritten
digit and street sign classifiers. After that, more realistic clean-label poisoning attacks that do not tamper with the label of the backdoored examples have emerged~\cite{shafahiPoisonFrogsTargeted2018, turner2019labelconsistent}.

In cybersecurity, a series of backdoor attacks have been designed via packet-level poisoning~\cite{holodnakBackdoorPoisoningEncrypted2022,ningTrojanFlowNeuralBackdoor2022}, feature-space poisoning~\cite{apruzzeseAddressingAdversarialAttacks2019,Li2018EPD}, and label flipping~\cite{Papadopoulos2021}. Furthermore, model explanability techniques have been used to develop clean-label backdoor attacks against malware classifiers on Windows PE files, Android applications and PDF files~\cite{severiExplanationGuidedBackdoorPoisoning2021}.
Recently, Severi et al.~\cite{severiPoisoningNetworkFlow2023} explored backdoor attacks on network traffic, where data dependency constraints, which inherently exist in traffic flows, make trigger mapping between feature and problem space particularly difficult. While their attacks were shown to be successful in various settings, no defense mechanism has been proposed.

\subsection{Mitigations against backdoor attacks}
\label{sec:rw_defenses}

Three main directions have emerged in the study of backdoor defenses. One approach has focused on developing techniques to obtain certifiable robustness guarantees to poisoning attacks, the second has looked at detecting and filtering out poisoned samples before training the model, while the third aims at directly purifying the poisoned model by unlearning the backdoor association.

Certified defenses to poisoning attacks aim to provide rigorous mathematical guarantees on model quality in the presence of poisons.  In some of the first work on certified defenses, Steinhardt et al.~\cite{Steinhardt2017Cert}  construct approximate upper
bounds on the loss using a defense based on 
outlier removal followed by empirical risk minimization. They show that an oracle defender who knows the true class mean is very powerful. However, a data-dependent defender that uses the empirical means performs poorly at filtering out attacks.  Deep Partition Aggregation (DPA) \cite{DPA2021} splits the training dataset into $k$ partitions, which limits the number of poisons that any one member of the ensemble can see.  Unfortunately, the number of partitions that are necessary can be high compared to the number of poisons considered.  To illustrate, using 250 partitions on CIFAR-10, DPA is able to certify 50\% of the dataset to just nine insertions/deletions applied to the training set. 

A related technique known as Deterministic Finite Aggregation \cite{DFA2022} provides modest improvements over DPA. In addition to the limited guarantees, a large number of partitions may not be practical in many security applications, as the number of malicious samples available is often small.    Randomized smoothing is a technique in which noise is added to the training data to attempt ``smooth-out'' any adversarial perturbations introduced by an attacker.  However, the guarantees that are enabled are fairly small.  Wang et al. \cite{Wang2020} are only able to certify 36\% of MNIST against an adversary that perturbs up to 2 pixels per image.

In the second category, a common detection strategy utilizes deep clustering, which consists in partitioning on the learned representation of the neural network. Chen et al.~\cite{chen2018detecting} propose \emph{activation clustering}, a defense method that runs 2-means clustering on the activations of the last hidden layer, then discards the smallest of the two clusters. 
Tran et al.~\cite{Tran2018SpectralSI} propose a defense based on \emph{spectral signatures}. This method computes outlier scores using singular value decomposition of the hidden layers, then removes the samples with top scores and re-trains.  An extension of this idea is used by the SPECTRE method \cite{Spectre}, which introduces the idea of estimating the mean and covariance matrices of the clean data, using robust statistics, and using these to whiten the data, prior to computing the singular value decomposition.

The third category includes various pruning and fine-tuning methods aimed at unlearning the backdoor pattern.
Liu et al.~\cite{liu2018fine-pruning} propose \emph{fine-pruning}, a strategy that attempts to disable backdoor behavior by eliminating less informative neurons prior to fine-tuning the model on a small subset of clean data.
Li et al.~\cite{Li2021NeuralAt} introduce \emph{neural attention distillation}, a purifying method that utilizes a teacher network to guide the fine-tuning of a backdoored student
network on a small subset of clean training data.
In a recent study, Heng et al.~\cite{Heng2023SEAM} employ \emph{selective amnesia} (SEAM), where catastrophic forgetting is induced by retraining the neural network on randomly labeled clean data; after both primary and backdoor tasks are forgotten, the randomized model is retrained on correctly labeled clean data to recover the primary task.
In \Cref{sec:seam} below, we evaluate a direct application of SEAM to the attacks we study, and show the drawbacks of that defensive approach in our scenario.
We note also that a similar approach, based on randomized unlearning and detection is proposed by Qi et al.~\cite{qiProactiveMLApproach2023}.

Other approaches in the literature involve detecting if a model is backdoored (ABS~\cite{ABS}, MNTD~\cite{MNTD}), detecting if a particular label has been under attack (Neural Cleanse~\cite{NeuralCleanse}), or performing topological analysis of activations at multiple layers in a neural network model (TED~\cite{TED}). In cybersecurity settings, the only defense we are aware of is Nested Training~\cite{NestedTraining, NestedTraining_Clean}, a data sanitization method that relies on an ensemble of multiple models, each based on different subsets of the training set. Nested Training has been applied to poisoning availability and backdoor attacks in network traffic and malware classifiers, but it incurs high cost and degradation in model performance.

\section{Problem Statement}
\label{sec:problem}
The goal of this paper is to study effective methods for defending against backdoor attacks in cybersecurity environments. We are focusing on feasible clean-label poisoning strategies that are not tailored to a specific model; instead, they are applicable to a variety of model architectures and data modalities used in security.

\subsection{Setting} 
\label{sec:motiv}
Compared to the task of protecting ML models designed for image classification or natural language processing (NLP) tasks, operating in a security sensitive environment presents both unique challenges and  opportunities. To start with, in this domain, an effective defensive approach has to be applicable to a wide variety of models, such as decision-tree based ensembles, including Random Forests and Gradient Boosting Trees, and neural network architectures. These models are state-of-the-art in many security applications, such as malicious domain classifiers~\cite{MADE} and malware classification~\cite{anderson2018ember}.

Moreover, the availability of a set of clean reference data points, a common assumption in backdoor defense research \cite{zhuSelectiveAmnesiaEfficient2023, qiProactiveMLApproach2023}, is not to be taken for granted.
It may indeed be relatively simple for an organization to obtain clean samples of images or natural language text, by scraping the Internet or leveraging large scale  crowd-sourcing systems such as Amazon Mechanical Turk to gather annotations or filter out potentially anomalous data points.
However, ensuring that cybersecurity data has not been tampered with requires potentially complex analyses by domain experts, which is typically a significantly more expensive process.

Conversely, the intrinsic constraints of cybersecurity data modalities also limit the attacker party. Thus, defensive measures designed for these models can rely on different assumptions, which are unlikely to hold in other domains. For instance, existing attacks~\cite{severiExplanationGuidedBackdoorPoisoning2021,severiPoisoningNetworkFlow2023} surfaced feature importance estimation as a key component in the design of an effective backdoor trigger. Triggers that do not leverage relevant features are usually not effective to induce the mis-classification at deployment time, as demonstrated by prior work~\cite{severiPoisoningNetworkFlow2023}. This intuition can, in turn, be used by a defensive mechanism to reduce the scope of the detection task in feature space.

Additionally, stealthy backdoor attacks on security classifiers tend to rely on clean-label attack formulations, where the poison is injected in benign data points, as the adversary often lacks the ability to interfere with the labeling mechanism.
This practice also has the added side effect of allowing the poisoning points to blend in among the usually large variety of benign points.
While the variance in benign data can be problematic from a defensive standpoint, this trend also implies that a defender can assume that the poisoned samples are a subset of the benign training data.

\subsection{Threat model}

Our work focuses on defending binary classification models designed for security applications, in particular tabular data classification on hand-crafted feature representations.
Therefore, the assumed \emph{adversarial objective} is to ensure the ability to arbitrarily trigger the output of \emph{benign} for an actually malicious data point chosen by the adversary.
In a practical setting this would correspond to the ability to evade the victim classifier, and allow an undesired action on the system, such as the execution of malware or the propagation of the malicious network flows.

Similarly, the \emph{defender objective} is to minimize, and potentially completely disrupt, the success rate of the attacker.
Parallel to this primary goal, the defender also strives to minimize the side effects of their defensive strategy on the performance of the model.
Of high concern is the false positive rate (FPR) --- the fraction of benign points classified as malicious --- of the model.
False positives are particularly expensive for security vendors, as they result in direct interference with the normal operations of their clients, and usually require a prompt investigation leading to overheads.
Therefore, an effective defensive mechanism should have minimal impact on the FPR of the defended model.

\subsubsection{Adversary capabilities}

We consider different adversaries with the range of capabilities studied in \cite{severiExplanationGuidedBackdoorPoisoning2021, severiPoisoningNetworkFlow2023}. In particular, the attacks assume the adversary to be knowledgeable of the feature representation used by the victim model. 
Moreover, the attacker is able to either query a version of the victim model, or use a surrogate to estimate feature importance values.
In all cases, the attackers are also characterized by the ability to introduce a small amount of poisoned samples in the training set, and a complete lack of control over the labeling step of the training pipeline.

\subsubsection{Defender capabilities}

In contrast to previous work, we do not assume that the adversary is in possession of a clean dataset distributed as the training set.
While this assumption can be true in some contexts (it is often rather easy to acquire, or verify, clean data for image classification tasks), we argue that it introduces large costs in the security domain, since manual analysis of the data points is a task that requires domain experts.
We do, however, assume the defender has the computational resources required to perform clustering on the training data and to train multiple models.
Differently from computer vision or natural language processing models, which are often billions of parameters large, security models tend to be smaller, often based on ensembles of decision trees and gradient boosting, making them relatively inexpensive to re-train.

\subsection{Challenges of existing defenses}
\label{sec:seam}

Most backdoor defenses proposed in the literature are not immediately applicable to cybersecurity. Security applications considered in this work
have unique characteristics, threat models, and assumptions that make applying existing defenses difficult:

\paragraph{Different data modalities and model architectures}
The majority of proposed mitigations have been  specifically designed for the computer vision domain where CNNs are state-of-the-art architectures~\cite{chen2018detecting,Tran2018SpectralSI,Spectre,TED}. Representative approaches for computer vision backdoor defenses are activation clustering~\cite{chen2018detecting}, spectral signatures~\cite{Tran2018SpectralSI}, and SPECTRE~\cite{Spectre} that perform outlier detection in the CNN representation space via clustering, SVD decomposition, or robust statistics.  Prior work~\cite{severiExplanationGuidedBackdoorPoisoning2021} tried to adapt these methods over the feature space  for defending against poisoning in malware classification, but they showed that these defenses are not effective when they do not operate in the model's representation space. 

\paragraph{Coarse-grained binary labels}
The typical cybersecurity task is to distinguish malicious and benign samples, resulting in a binary classification problem. Techniques that look for shortcuts between classes in latent space and aim to identify a target attack label, such as Neural Cleanse~\cite{NeuralCleanse} and ABS~\cite{ABS}, require finer-grained labeling of the training data and thus cannot be readily adapted to our setting.

\paragraph{Hard constraints on false positives} 
Models trained for cybersecurity tasks have hard constraints on false positives to be deployed in production~\cite{MADE}. This requirement rules out the application of certified defenses~\cite{Steinhardt2017Cert,Wang2020,DPA2021,DFA2022} which largely degrade model utility. 

\paragraph{Applicable defenses}
Among the defenses surveyed in Section~\ref{sec:rw_defenses}, the only category that can be applied in cybersecurity settings is the pruning or fine-tuning
methods aimed at unlearning the backdoor pattern~\cite{liu2018fine-pruning,Li2021NeuralAt,Heng2023SEAM,qiProactiveMLApproach2023}. As discussed, all these approaches require the availability of a clean dataset, an assumption not easily satisfied in our setting. Nevertheless, we select a recent representative technique in this category, Selective  Amnesia~\cite{Heng2023SEAM}, and show its limitations when applied in this domain. 

\subsection*{Limitations of Selective Amnesia}

\begin{figure}
\includegraphics[width=8cm]{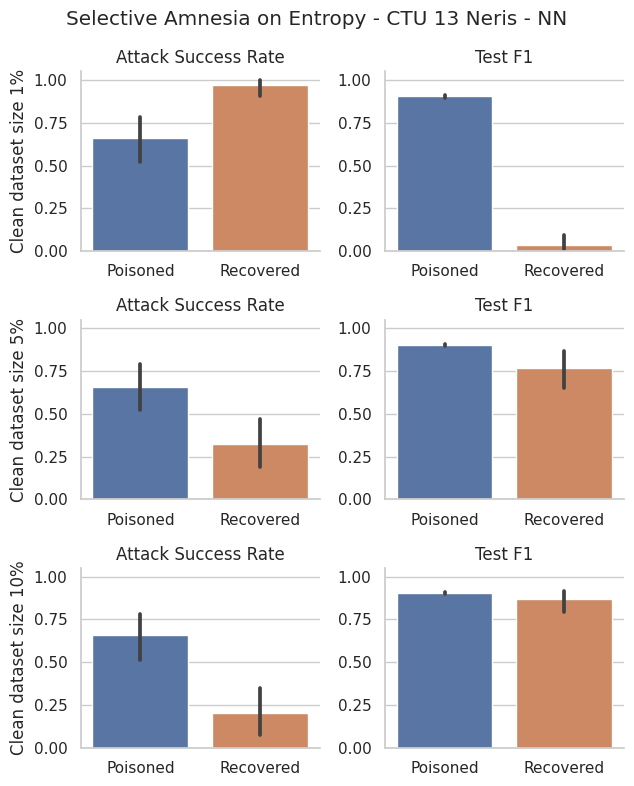}
\caption{Selective Amnesia defense applied to the attack against the CTU-13 Neris botnet classifier. The plots compare in attack success rates before and after recovery, and the F1 score on test data, for different sizes of the clean dataset. Attack run with Entropy feature selection.}
\Description{Selective Amnesia defense applied to the attack against the CTU-13 Neris botnet classifier. The plots compare in attack success rates before and after recovery, and the F1 score on test data, for different sizes of the clean dataset. Attack run with Entropy feature selection.}
\label{fig:seam_ctu_entropy}
\end{figure}

We adapt the Selective Amnesia (SEAM) defense~\cite{zhuSelectiveAmnesiaEfficient2023} on one of our security scenarios, the CTU-13 Botnet detection task. 
Conceptually, SEAM works by forcing the model to forget the association with the trigger by inducing catastrophic forgetting through fine-tuning on randomly assigned labels different from the ground truth.
Once the model has forgotten both the primary classification task, and the backdoor association, SEAM proceeds to fine-tune the model on a held out clean dataset until the accuracy of the model is restored.

Selective amnesia was designed for multi-class classification in vision tasks, therefore we had to re-implement the method adapting it for our binary classification task on security data.
Moreover, SEAM requires the ability to fine-tune a model on both mis-labeled points during the forgetting phase, and on clean data during the recovery phase.
Thus, we only tested it on our neural network classifier, as the process of fine tuning different types of classifiers (e.g., Random Forests, Gradient Boosting Trees) is not well defined.

\Cref{fig:seam_ctu_entropy} reports the results of SEAM applied to clean-label backdoor attacks on security data, for different sizes --- expressed as percentage of the training set --- of the held-out clean dataset.
We observe that SEAM's success as a mitigation technique is strongly dependent on the size of the clean dataset: 1\% recover set size is insufficient to prevent the attack, while 5\% clean dataset size decreases the attack success rate from approximately 0.65 to 0.30. The defense becomes more effective, but still does not completely thwart the attack, as the clean dataset size reaches 10\%. 
Since, for cybersecurity tasks, acquiring and validating large volumes of data is markedly more expensive than in other domains such as NLP or computer vision, we believe that there is a need for a mitigation approach that doesn't directly rely on access to clean data.

\begin{figure*}[th]
    \centering
    \includegraphics[width=0.8\linewidth]{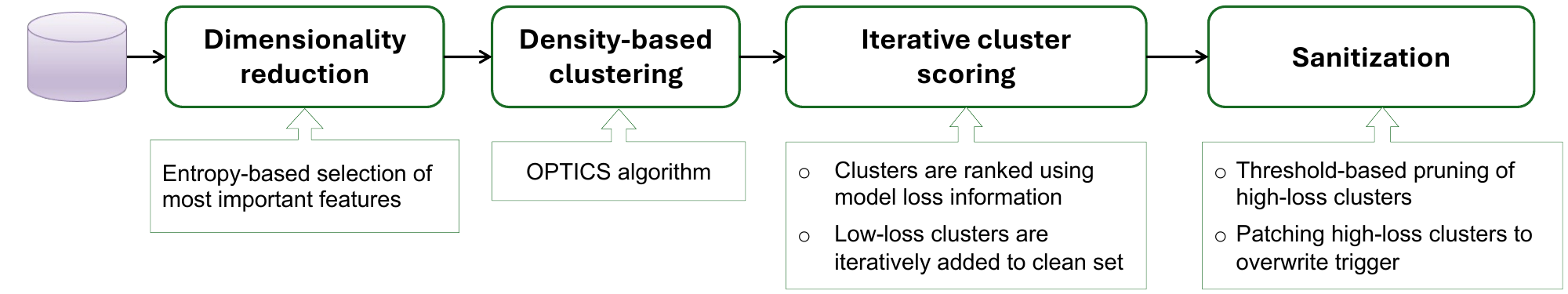}
    \caption{Pipeline of our defense strategy.}
    \Description{Pipeline for our defense strategy.}
    \label{fig:pipeline}
\end{figure*}

\section{Defense strategy}

Guided by the considerations presented in \Cref{sec:problem}, we develop a defensive strategy aimed at protecting models designed for security classification tasks against stealthy clean-label poisoning attacks. 
Our procedure for sanitizing the training dataset proceeds in several stages, outlined in Figure~\ref{fig:pipeline}. A more detailed pseudo-code of our strategy is presented in Algorithm~\ref{alg:method}. Before detailing each stage in the defense pipeline, we provide some key insights that enable us to address the limitations of prior methods.

\noindent {\bf Dimensionality reduction}: First, motivated by the observation that most clean-label poisoning attacks in cybersecurity leverage important features~\cite{severiExplanationGuidedBackdoorPoisoning2021,severiPoisoningNetworkFlow2023}, we perform dimensionality reduction by selecting the most relevant features for classification. For this stage, we use an entropy metric computed on a decision tree model fitted to a subset of the data for identifying the top contributing features, and perform our subsequent analysis in this reduced space. 

\noindent {\bf Density-based clustering}: In the second stage, we perform clustering of samples with the benign label in the reduced space, to identity the clean-label poisoned samples. Our main insight is that poisoned samples lay in a different subspace of the benignly labeled samples and they will cluster together as they have similar values in the trigger.  
We leverage density-based clustering to partition the training data based on similarity in feature space. In contrast to centroid-based methods, density-based clustering determines the number of clusters dynamically, can detect irregularly-shaped clusters, and can handle both dense and sparse regions. 

\noindent {\bf Iterative cluster scoring:} Due to the stealthy nature of poisoning attacks, poisoned clusters will be small in size. We can thus assume that the largest cluster (usually including at least $50\%$ of the training set) consists of clean benign samples, an assumption we will validate and confirm experimentally. Hence, we select the largest cluster to initialize the clean set and conduct an iterative process where we gradually select clean clusters and re-train the model. The main challenge is identifying the poisoned clusters during this iterative process, so that we can exclude them from model training. After adding a set of clean clusters to the model in a particular iteration, we evaluate the model loss on the remaining clusters. Intuitively, clusters with lowest loss are closer to the (clean) training set. With this insight, we progressively grow the clean set by adding the lowest-loss clusters at each iteration. 

\noindent {\bf Sanitization of high-loss clusters}: Lastly, we employ a data sanitization step to either filter or patch the high-loss clusters to protect the model against poisoning. For the filtering strategy, we either stop the iterative process after including a fixed percentage (e.g., 80\%) of clusters in the training set, or select a subset of clusters to exclude via loss analysis. For the patching strategy, we include all clusters in training, but apply a patch to the most relevant features in the highest loss clusters. 

We now describe in details each stage of the defense pipeline. 

\subsection{Dimensionality reduction}

Clustering high-dimensional data is affected by the curse of dimensionality -- as dimensionality increases, data points become more dissimilar or farther from each other~\cite{Steinbach2004}. Therefore, before running a clustering technique, we first reduce the dimensionality of the data points to a feature subset $\mathcal{F}$.
Here we leverage the following insight about the attack process: strong backdoor attacks choose features with high impact over a model's decisions~\cite{severiExplanationGuidedBackdoorPoisoning2021,severiPoisoningNetworkFlow2023}. Hence, we reduce the data set to the top $\mathcal{|F|}$ most important features.

We would like to compute feature importance in a model-agnostic fashion. In tree-based models, entropy is used to decide which feature to split on at each step in building the tree. The lower the entropy, the more beneficial the chosen feature is. We employ a similar technique by mapping a surrogate decision-tree model on our data and computing entropy-based feature importance to rank the features and select the top most relevant ones. Specifically, in our experiments, we reduce the network traffic and binary files datasets from approximately 1100 and 2300 features, respectively, to top 4 and 16 most important features.

\subsection{Clustering}
Our defense relies on a clustering procedure to isolate the compromised training points. 
While previous work such as Chen et al.~\cite{chen2018detecting} is tailored to deep neural network models --- they apply clustering on the last hidden layer representation of the neural network model to distinguish between poisonous and legitimate data --- we design a model-agnostic strategy. 

We perform density-based clustering operating in the reduced feature space $\mathcal{F}$, ensuring a broader applicability to multiple classification models. We use OPTICS (Ordering Points To Identify the Clustering Structure)~\cite{ankerst1999optics} to find high-density cores and then expand clusters from them. OPTICS automatically identifies the number of clusters, $k$, present in the data and works better than the commonly used density-based method DBSCAN~\cite{dbscan1996} on clusters with different densities, a characteristic that is particularly useful for our defense.

This approach reliably isolates the poisoned points into a few clusters.
However, the defender still does not know which clusters are corrupted. 
Therefore, the next step of our defense consists of designing scoring and filtering systems that can leverage the information advantage gained through clustering.

\subsection{Cluster loss analysis}
\label{sec:intuition}

Prior to discussing our cluster scoring methodology, we provide some intuition as to how model loss behaves on the clustered dataset.  
Our strategy takes into account two defining characteristics of stealthy backdoor attacks. First, the adversary strives to minimize the attack footprint, i.e., the number of attack points introduced in the training data.
Therefore, we can conclude that poisoned data points generally reside in relatively small clusters, while larger clusters are mostly clean. Second, in an attempt to camouflage the poisons as benign, the adversary inserts the backdoor into data points belonging to the \emph{benign} class. Based on this insight, we compute the clusters only over $D_{y=0}$ -- target-class (benign) data, while the non-target (malicious) training points $D_{y=1}$ can be assumed clean (i.e., not containing the backdoor), and used to train surrogate models.

\begin{figure}
    \centering
    \includegraphics[width=0.75\linewidth]{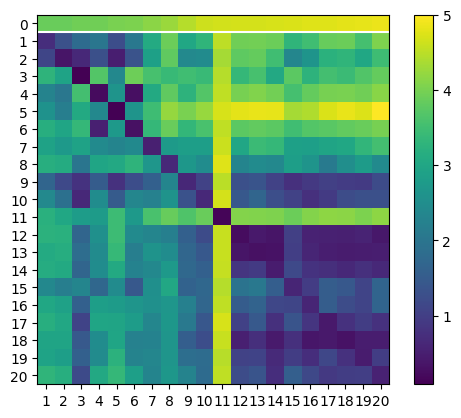}
    \caption{Row 0: Log-loss of model trained on $C_0 \bigcup D_{y=1}$ and evaluated on clusters $C_j$.  Rows 1-20: Log-loss of model trained on $C_0 \bigcup C_i \bigcup D_{y=1}$ and evaluated on clusters $C_j$.  Note that cluster 11 consists of poisoned data and the remainder contain only clean data. Experiment on CTU-13, gradient boosting classifier, attack run with entropy feature selection.}
    \Description{Row 0: Log-loss of model trained on $C_0 \bigcup D_{y=1}$ and evaluated on clusters $C_j$.  Rows 1-20: Log-loss of model trained on $C_0 \bigcup C_i \bigcup D_{y=1}$ and evaluated on clusters $C_j$.  Note that cluster 11 consists of poisoned data and the remainder contain only clean data. Experiment on CTU-13, gradient boosting classifier, attack run with entropy feature selection.}
    \label{fig:heatmap}
\end{figure}

We anticipate that models trained on clean data will exhibit very high loss on clusters containing poisoned data and models trained on poisoned data will exhibit very low loss on clusters containing poisoned data.  Assuming that the largest cluster (which we denote as $C_0$) is likely to be clean, we consider comparing the loss of a model on cluster $j$, when the models is trained on $C_0 \bigcup D_{y=1}$ versus when it is trained on  $C_0 \bigcup C_i \bigcup D_{y=1}$, for each cluster $i$.  If cluster $j$ contains poisons, we anticipate that the change in loss will be small, apart from those few clusters $i$ that also contain poisons.  An example of carrying out this procedure on a poisoned dataset that the defender has clustered is shown in Figure \ref{fig:heatmap}, where cluster 11 consists of poisoned data and the remaining clusters consist of clean data.  The twenty clusters shown (out of about 700) are those with the highest loss for a model trained on $C_0 \bigcup D_{y=1}$.  Note that the model has high loss on cluster eleven when trained on $C_0 \bigcup D_{y=1}$ and exhibits almost no improvement except when the model is trained on $C_{11}$.  We observed that this method was often effective at identifying clusters in our experiments, but it carries a high computational cost as a model must be trained for each cluster. 
 In some of our experiments, we observed more than 1000 clusters, which motivated us to search for a more efficient method to identify the poisoned clusters.

\subsection{Iterative cluster scoring}
\label{sec:loss_analysis}

We develop an iterative scoring procedure where clean clusters are progressively identified and added to the clean data set.

Our cluster scoring procedure is presented in Algorithm~\ref{alg:method}, lines 9-27. We start by selecting the largest cluster $C_0$ out of the entire set of clusters $C$ identified by OPTICS, and merging it with the malicious points, to generate a temporary (small) clean training set, $D_{clean} \leftarrow C_0 \bigcup D_{y=1}$.
Next, we train a clean model $f$ on $D_{clean}$, and evaluate $f$'s average loss on each remaining cluster $C_i \in C \setminus C_0$. We rank the clusters $C_i$ by loss, bearing in mind that, typically, the lower the loss, the closer the cluster is to the training data $D_{clean}$.

Having defined a fixed window size $w$ -- we use $w = 5\%$ of the clusters -- we construct an iterative filtering process as follows. 
We take the $w$ of the clusters with the lowest average loss score on the clean model, and add their data to the clean training set $D_{train}$. 
Then we re-train the surrogate model $f$ on this new dataset and score the remaining clusters.
This process iteratively pushes the clusters that are less similar to the data assumed to be clean to the bottom of the list, therefore isolating the clusters containing poisoned data points.

At this point the defender could opt for a fixed threshold filtering strategy, by stopping the iterative training and scoring process after a fixed percentage of clusters has been added to $D_{clean}$.
In our experiment, we set this fixed threshold to 80\% of the clusters, as we empirically find it to be quite effective.
However, based on the intuition developed in \Cref{sec:intuition}, we expect the poisoned clusters to affect the loss of the model in a very specific way when added to the training set.
Therefore, we can proceed to look for those clusters for which we observe significant changes between the loss of the model computed on the cluster before it being introduced in $D_{clean}$ and after the model trained on it.
By measuring these deltas in the loss for each cluster, we can look for anomalous clusters using a standard statistical approach, the Z-score\footnote{\url{https://en.wikipedia.org/wiki/Standard_score}}.

In practice, we accumulate the loss deltas, $\Delta_{i} = loss(f_{current}, C_i) - loss(f_{previous}, C_i)$,  for each cluster $i$ when it is included in $D_{clean}$ up to the 80\% threshold. 
Then we use the recorded values of $\Delta$ to compute mean $\mu$ and standard deviation $\sigma$.
Then we compute $z = \frac{x - \mu}{\sigma} $ for all the clusters, define a threshold $z_t = 2 \sigma$, and apply our sanitization procedure to those clusters with score $z <= z_t$.

\subsection{Sanitization of high-loss clusters}

The iterative process described above effectively isolates the poisoned points. 
The final step of our defense consists in dealing with the isolated clusters. 
We identify two main remediation procedures: 
\begin{enumerate*}[label=(\roman*)]
  \item discarding clusters that have not been considered clean;
  \item patching the data points belonging to clusters that that have not been considered clean. 
\end{enumerate*}
In \Cref{sec:eval} below, we show the effects of applying both strategies to the suspicious clusters identified by each of the two methods, i.e., (a) clusters are considered suspicious after a fixed number of iterations, and (b) the pre-clustering loss deltas are analyzed to surface suspicious clusters.

The first strategy simply removes clusters that are considered suspicious from the training set of the final model, thus preventing the poisoned points from corrupting the model.
This procedure has the advantage of being easy to implement, but, depending on the number of clusters removed, may damage the utility of the model, especially on uncommon test points.

The second approach aims at preserving as much utility of the model as possible by still extracting information from points belonging to suspicious clusters, while at the same time minimizing the effect of the attack.
It consists in applying a patch over the subspace of the $\mathcal{P}$ most important features to each data point in the suspect clusters, while maintaining unaltered the rest of the vector.
Note that in general $|\mathcal{P}| \geq |\mathcal{F}|$, as the defender can be conservative and patch a larger portion of the feature space than the one used for clustering.

In our implementation, we randomly sample values for the patched features from the set of points in $D_{clean}$, when $D_{clean}$ includes 80\% of the clusters.
With the continuous evolution of generative modeling for tabular data -- for instance with applications of diffusion models to tabular modalities --, however, the defender could design patching mechanisms that could potentially lead to even larger gains in model utility.
We leave the exploration of this interesting research thread for future work.

\begin{algorithm}

\SetKwRepeat{Do}{do}{while}
\SetKwFunction{Algo}{Defense}
\SetKwProg{Fn}{Procedure}{:}{}

\KwData{
$D$: training data set in feature space; \newline
$D_{y=0}$: the subset of $D$ labeled benign (examples may include trigger); \newline
$D_{y=1}$: the subset of $D$ labeled malicious }

\Fn{\Algo{$D$}} {

\textcolor{gray}{// reduce the data set to the most important features}

$\mathcal{F} \leftarrow 
\textsc{dimensionality\_reduction}(D_{y=0})$

\textcolor{gray}{// partition the data set into clusters}

$C \leftarrow 
\textsc{density\_based\_clustering}(\mathcal{F})$

\textcolor{gray}{// initialize the clean set with the largest cluster}

$C_0 \leftarrow max(C)$

\textcolor{gray}{// use model loss to isolate clusters that contain poisons}

\Do {$C_{0} \neq D || stop\_condition$} {

\textcolor{gray}{// current clean data (both benign and malicious)}
    
$D_{clean} \leftarrow C_0 \bigcup D_{y=1}$

\textcolor{gray}{// train a new clean model $f$ on $D_{clean}$}
    
$f \leftarrow \textsc{train}(D_{clean})$

\textcolor{gray}{// remaining clusters to be scored and filtered}

$C_r \leftarrow C \setminus C_0$

\textcolor{gray}{// dictionary of loss per cluster}

$\mathcal{L} = \{\}$

\textcolor{gray}{//evaluate loss for each remaining cluster}

\For{$i \in range(|C_r|)$}{

    \textcolor{gray}{// evaluate the average loss over all data points}
    
    \textcolor{gray}{// in cluster $C_i$ using the current clean model $f$}
    
    $\mathcal{L}[i] = \textsc{compute\_loss}(f, C_i)$
}

\textcolor{gray}{// add cluster(s) with lowest loss to $C_0$ and repeat}

$C_l \leftarrow \text{lowest loss cluster(s) from}\mathcal{L}$

$C_0 \leftarrow C_0 \bigcup C_l$
}

\textcolor{gray}{// consider remaining clusters as suspicious}

$C_r \leftarrow C \setminus C_0$

$ C_r \leftarrow \textsc{patch\_or\_discard}(C_r)$

\textcolor{gray}{// train the final purified model}
    
$D_{clean} \leftarrow C_0 \bigcup C_r \bigcup D_{y=1}$

$f \leftarrow \textsc{train}(D_{clean})$

$\texttt{{\bf return }} f$
}

\caption{Mitigation Procedure}
\label{alg:method}
\end{algorithm}

\section{Evaluation}
\label{sec:eval}

In this section we report the results obtained in our experimental evaluation.
We start with describing the experimental setup, and then we explore the settings of network flows and binary files classification.

\subsection{Experimental setup}
We report here the results of evaluating our defensive approach on the attacks proposed in \cite{severiExplanationGuidedBackdoorPoisoning2021,severiPoisoningNetworkFlow2023}.
We use the experimental conditions reported in the papers, and, following the original works, we test our defense on different model types, such as neural networks, gradient boosting trees, and LightGBM classifiers.

For generality, we apply our defense to two types of data: network flow traffic and binary files.
In particular, we run the network data poisoning attack on the CTU-13 dataset for the Neris botnet detection task \cite{garciaEmpiricalComparisonBotnet2014}, and the executable poisoning on the EMBER malicious Windows Portable Executable file dataset \cite{anderson2018ember}.
Due to the large number of experiments we run, to speed up the experimental phase, we subset the EMBER dataset selecting randomly 10\% of the original 600,000 labeled data points, preserving the 50\% class split.

\subsubsection{Feature representation.} 
The network traffic datasets consist of NetFlow data, i.e., connection events that have been extracted from packet-level PCAP files using the Zeek monitoring tool~\cite{zeek}. 
The classifier uses a subset of the Zeek log fields, which can be more effectively associated with intrusion, such as: IP address, port, number of packets and payload bytes (for both originator and responder), as well as protocol, service, timestamp, duration, and connection state.

A sequence of this network log data (e.g., 30s time window) is mapped to a feature-space data point using statistic-based aggregation techniques. 
Statistical features are commonly used to analyse large volumes of traffic and detect suspicious network activity~\cite{mooreDiscriminatorsUseFlowbased2005,yangFeatureExtractionNovelty2021,Burkhart2010FeatureStatistics}.  
In line with previous work~\cite{ongunPORTFILERPortLevelNetwork2021,severiPoisoningNetworkFlow2023}, the statistical features we use are aggregated by time window, internal IP and destination port and include several count-based metrics such as: counts of connections per transport protocol and connection state, counts of distinct external IPs communicating with each internal IP (as either originator or responder).
Sum, min, and max statistics of traffic volume (packets/bytes) are also included.

For malware binary files, the classifier operates in the feature space provided by the EMBER dataset~\cite{anderson2018ember}.
A feature-space data point contains information regarding both metadata (e.g., image, linker and operating system version), as well as statistical information (e.g., number of URLs and file system paths, and number of read, write and execute sections).

The full list of features is presented in~\Cref{sec:features}.

\subsubsection{Attacks.} We evaluate our defense on the backdoor attacks proposed by Severi et al.~\cite{severiPoisoningNetworkFlow2023} against network flow classifiers. The attacks explore both the entropy-based feature selection -- through a surrogate decision tree -- method, and SHAP~\cite{SHAP}, a game-theory inspired explanation technique that queries the model to compute feature relevance coefficients. 
Two types of triggers are considered: 1) full trigger, a contiguous subset of log connections where the backdoor pattern is embedded in the most important features, and 2) generated trigger, a stealthier variant where poisons resemble benign data. The generative trigger is constructed using Bayesian network models with the goal of preserving data dependencies in the network traffic, while attempting to blend the poisoned data with the underlying training distribution.

In addition, we evaluate our defense against backdoor attacks that target malware classifiers on Windows PE (Portable Executable) files. These attacks are presented in~\cite{severiExplanationGuidedBackdoorPoisoning2021}. They leverage SHAP-based explainability methods to guide the attacker towards the most informative features used in the classification process. 
We  have run these attacks using their publicly available source code from GitHub.

\subsubsection{Evaluation metrics.}

In our experiments we are interested in a few key metrics.
First we keep track of the \emph{attack success rate} (ASR), that is defined in \cite{severiExplanationGuidedBackdoorPoisoning2021, severiPoisoningNetworkFlow2023} as the percentage of backdoored test points, which would otherwise be correctly classified as malicious by a clean model, which are misclassified as benign by the poisoned model.
We are also interested in recording the fraction of poisoning points that manages to slip through our defenses, so we report the \emph{fraction of poisoning points} that ends up being included in $D_{clean}$.

In addition to these metrics oriented at measuring the effect of the mitigation on the attack, we also want to measure the side effects of our mitigation on model utility.
Therefore, we report the measured \emph{F1 score} and \emph{false positive rate} (FPR) of the defended model on clean test data.

\subsection{Evaluation on network traffic classification}

\begin{figure}[t]
\centering
\includegraphics[width=0.95\linewidth]{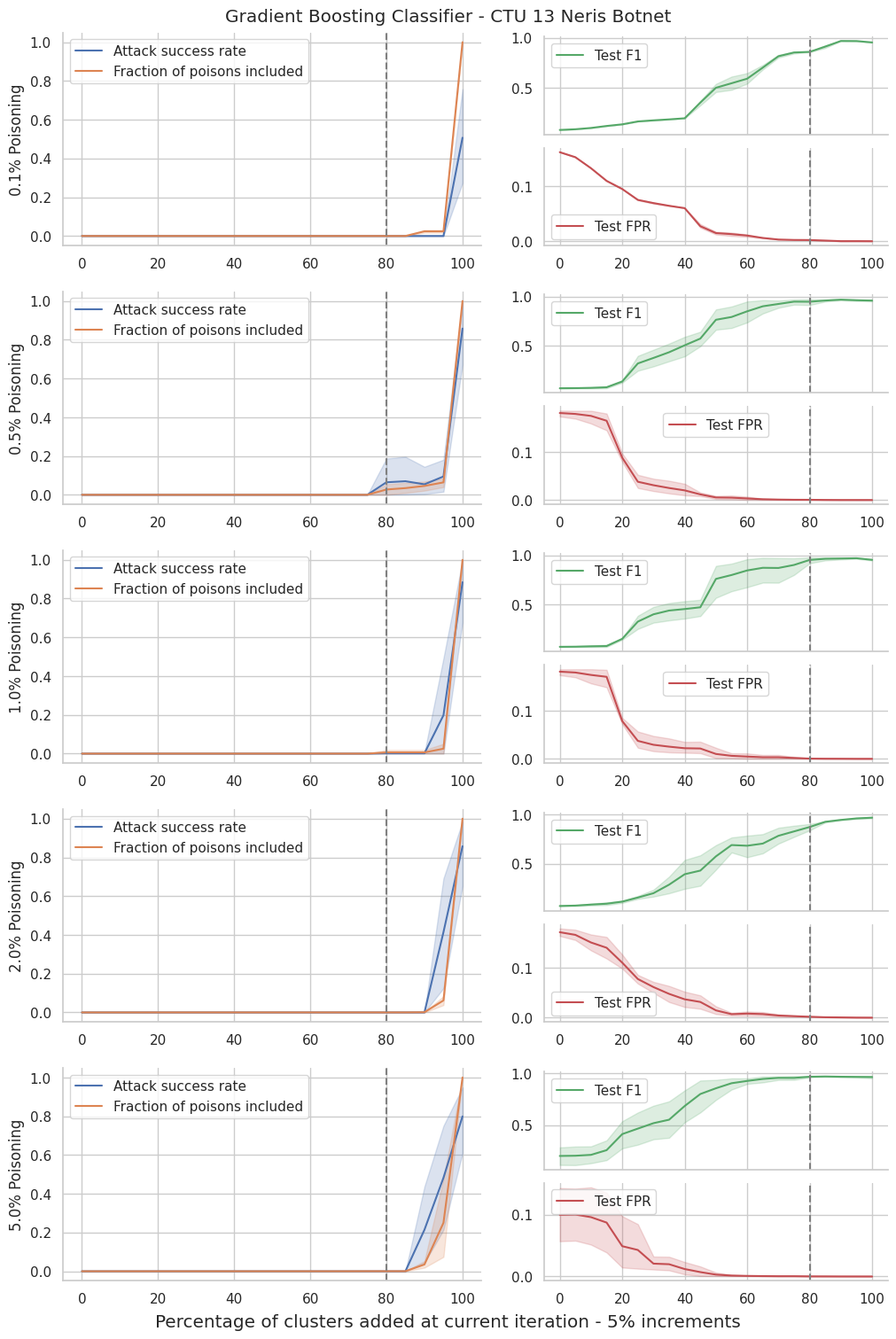}
\caption{Iterative scoring on the CTU-13 botnet classification task for the \emph{gradient boosting} model. The plot shows average metrics for a set of experiments: SHAP and Entropy attacker feature selection, for the Full trigger attack, at 5 different poisoning rates.}
\Description{Iterative scoring on the CTU-13 botnet classification task for the \emph{gradient boosting} model. The plot shows average metrics for a set of experiments: SHAP and Entropy attacker feature selection, for the Full trigger attack, at 5 different poisoning rates.}
\label{fig:gb_ctu13_iterative}
\vspace{-2pt}
\end{figure}

\begin{figure}[t]
\centering
\includegraphics[width=0.95\linewidth]{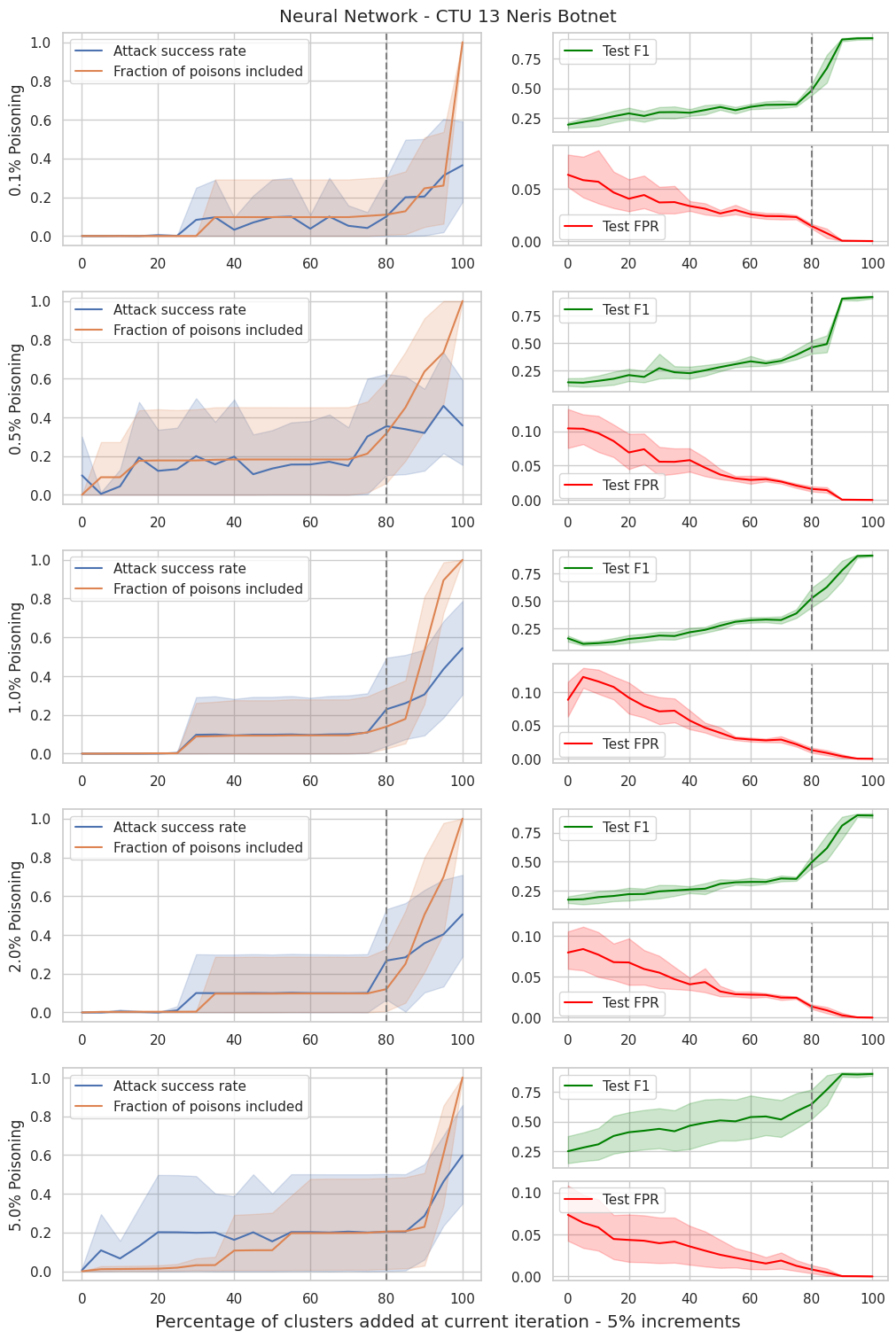}
\caption{Iterative scoring on the CTU-13 botnet classification task for the \emph{neural network} model. The plot shows average metrics for a set of experiments: SHAP and Entropy attacker feature selection, for the Full trigger attack, at 5 different poisoning rates.}
\Description{Iterative scoring on the CTU-13 botnet classification task for the \emph{neural network} model. The plot shows average metrics for a set of experiments: SHAP and Entropy attacker feature selection, for the Full trigger attack, at 5 different poisoning rates.}
\label{fig:nn_ctu13_iterative}
\vspace{-2pt}
\end{figure}

\Cref{fig:gb_ctu13_iterative} and \Cref{fig:nn_ctu13_iterative} show the dynamics of the metrics we tracked across the iterations, assuming a window size of 5\% of the clusters.
The reported metrics are aggregated through averaging over multiple experiments, with two different attack feature selection strategies (Entropy and SHAP), for the Full trigger attack. The results are shown at 5 different poisoning percentages, and each experiment is repeated 5 times with different random seeds.

\subsubsection{Fixed threshold filtering}

We start by considering the fixed threshold sanitization baseline.
From the figures we can observe that in the case of the full trigger attack, for both gradient boosting model and neural network, interrupting the training when 80\% of the clusters has been added to the training set represents a good baseline heuristic that filters out the vast majority of poisoning points. 
We also note that the poisoning fraction used by the attacker has little influence on the effectiveness of this procedure.

The generated trigger, on the other hand, is much stealthier, and a larger fraction of poisoning points manages to pass through the filtering process at these pre-defined thresholds, as highlighted in \Cref{fig:gb_ctu13_iterative_gen} in \Cref{app:iterative-scoring}. 
However, given that this attack is designed for stealthiness rather than effectiveness, the attack success rate is limited even if a fraction of the contaminants ends up in the training data.
Finally, we note here that, as initially observed in \cite{severiPoisoningNetworkFlow2023}, the generated trigger for the neural network model has a strong adversarial example effect at inference time. 
That is, the trigger pattern itself induces the classifier to misclassify the point, even if no poisoning attack took place.
Therefore, this case falls outside the scope of our defense, which is targeted at countering backdoor poisoning attacks.

While effective at removing the poisons (ASR is ranging from 0.0\% to 6.45\%), this baseline remediation strategy may reduce the utility of the models as a side-effect of discarding the entire clusters below the threshold (including their clean data). The F1 and FPR utility metrics reported in \Cref{tab:numbers_at_t}, \Cref{app:numbers_at_t} average from 0.02\% to 0.21\% for FPR, and 86\% to 97\% for F1.

\subsubsection{Patching.}

\begin{table}[t]

\caption{Comparison of patching and filtering sanitization approaches at fixed threshold = 80\%. Gradient boosting model on the CTU-13 dataset. Also showing the Base ASR value for the undefended attack. Results are averages of 5 runs on different random seeds, for two attack strategies Entropy and SHAP.}
\label{tab:patching80-ctu}

\begin{adjustbox}{width=0.98\linewidth}

\begin{tabular}{ll | r | rrr}
{\bf Sanitization} & {\bf Poisoning} & {\bf Base ASR} & {\bf ASR} & {\bf Test FPR} & {\bf Test F1} \\ \hline

\multirow{5}{*}{Filtering} & 0.1\%  & 50.70\% &  0.00\% &        0.21\% &      85.93\% \\
          & 0.5\% & 85.80\% &  6.45\% &        0.04\% &      94.84\% \\
          & 1\% & 88.45\% & 0.15\% &        0.04\% &      95.43\% \\
          & 2\% & 85.75\% & 0.00\% &        0.18\% &      87.13\% \\
          & 5\% & 79.90\% & 0.00\% &        0.02\% &      96.50\% \\
\hline
\multirow{5}{*}{Patching} & 0.1\% & 50.70\% &  0.00\% &        0.08\% &      90.61\% \\
          & 0.5\% & 85.80\% & 10.95\% &        0.05\% &      93.19\% \\
          & 1\% & 88.45\% &  9.75\% &        0.03\% &      93.37\% \\
          & 2\% & 85.75\% &  9.30\% &        0.05\% &      92.64\% \\
          & 5\% & 79.90\% & 10.95\% &        0.00\% &      95.27\% \\

\end{tabular}
\end{adjustbox}
\end{table}

The patching-based sanitization strategy addresses the degradation in utility metrics.
Considering the same threshold at 80\%, we can directly compare the effects of patching to filtering for the experiment with the gradient boosting model on CTU-13. 
\Cref{tab:patching80-ctu} shows that using patching generally leads to higher average values of the F1 score on test data (91\% - 95\%), and a lower false positive rate (0\% - 0.08\%).
On the other hand, as expected, patching is slightly less effective than complete filtering in reducing the attack success (0\% - 11\% ASR).
Therefore, the defender can adopt the patching approach if they want to trade-off some defensive effectiveness for reduced degradation in model utility.

\subsubsection{Loss analysis and sanitization}

We performed the loss deltas analysis outlined in \Cref{sec:loss_analysis} applying it to the experiments on the CTU-13 dataset with the gradient boosting model.
For this experiment we use a threshold Z-score of 2, that is, we mark as suspicious any cluster whose loss delta is lower than 2 times the standard deviation of the observed loss deltas during the initial training iterations, up to 80\%.
The results of both filtering and patching sanitization strategies are reported in \Cref{tab:patching-loss-280-ctu}, \Cref{app:loss_analysis}.

This approach leads to a significant improvement of the F1 and FPR scores, especially for the filtering sanitization.
In this case, we observe up to a ~0.2\% (95\% relative improvement) decrement in FPR and up to ~11\% increase (12.5\% relative improvement) in the F1 score on the test set.
However, the attack success remains relatively high, and therefore fixed threshold filtering may be preferable in this scenario.
We note here that the defender can always trade off smaller improvements in F1 and FPR scores for a stronger protection from the attack, by increasing the value of the threshold, for instance using $z_t=1\sigma$.

\subsection{Evaluation on malware classification}

\begin{figure}[t]
\centering
\includegraphics[width=0.95\linewidth]{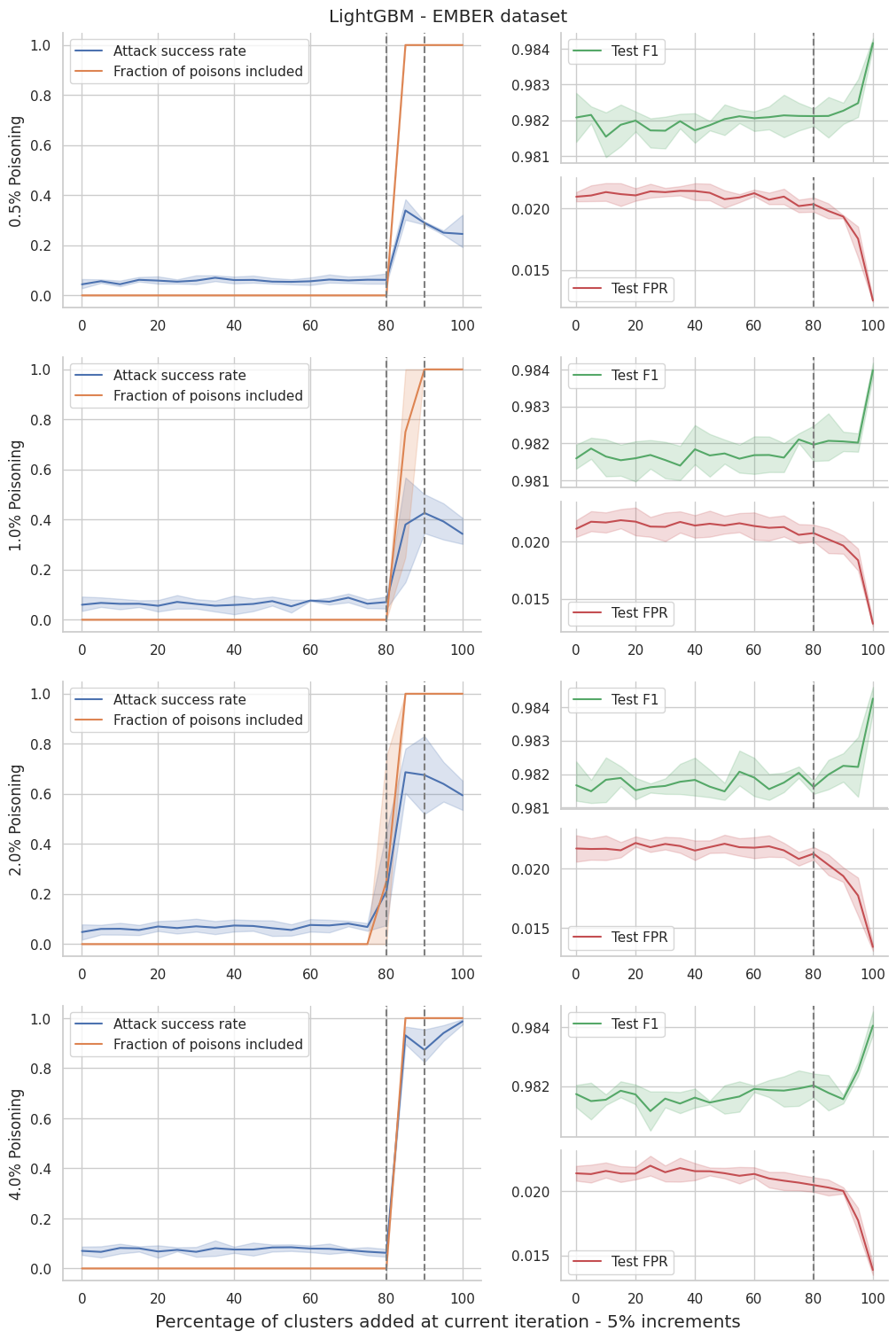}

\caption{Iterative scoring on the EMBER malware classification task for the \emph{LightGBM} model. The plot shows average metrics for a set of experiments: MinPopulation and CountAbsSHAP attack strategies, at 4 different poisoning rates.}
\Description{Iterative scoring on the EMBER malware classification task for the \emph{LightGBM} model. The plot shows average metrics for a set of experiments: MinPopulation and CountAbsSHAP attack strategies, at 4 different poisoning rates. }
\label{fig:lgb_ember_iterative}
\vspace{-2pt}
\end{figure}

\begin{table}[th]

\caption{Comparison of patching and filtering sanitization approaches at fixed threshold = 80\%. LightGBM model on the EMBER dataset. Also showing the Base ASR value for the undefended attack. Results are averages of two runs on different random seeds, for the two Independent attack strategies.}
\label{tab:patching80-ember}

\begin{adjustbox}{width=0.98\linewidth}

\begin{tabular}{ll | c | rrr}

{\bf Sanitization} & {\bf Poisoning} & {\bf Base ASR} & {\bf ASR} & {\bf Test FPR} & {\bf Test F1} \\ \hline

\multirow[c]{4}{*}{Filtering} & 0.5\% & 24.53\% & 6.19\% & 2.03\% & 98.21\% \\
 & 1\% & 34.27\% & 7.04\% & 2.08\% & 98.20\% \\
 & 2\% & 59.42\% & 20.98\% & 2.13\% & 98.16\% \\
 & 4\% & 98.67\% & 6.22\% & 2.05\% & 98.20\% \\
 \hline
\multirow[c]{4}{*}{Patching} & 0.5\% & 24.53\% & 25.66\% & 1.65\% & 98.31\% \\
 & 1\% & 34.27\% & 25.13\% & 1.64\% & 98.29\% \\
 & 2\% & 59.42\% & 68.18\% & 1.64\% & 98.34\% \\

\end{tabular}

\end{adjustbox}
\end{table}

We observe similar trends in our experiments on the malware classification task, as summarized in \Cref{fig:lgb_ember_iterative}.
These experiments were run attacking the LightGBM classifier, proposed with the EMBER dataset, using the two strongest attack strategies, based on independent feature and value selection: CountAbsSHAP and MinPopulation.
Each experiment was run twice with different random seeds, and we tested 4 different poisoning rates from 0.5\% to 4\%.

\subsubsection{Fixed threshold filtering}

In this scenario we observe a quicker rise in the fraction of poisoning points included in $D_{clean}$ after the 80\% threshold.
Notwithstanding, in general the fixed threshold heuristic is still quite effective in thwarting the attack.
Compared to the experiments on the network classification task, the changes in both F1 and FPR scores are sharper at higher percentages of included clusters, even if the absolute values of the decrements is generally smaller. We report 2\% FPR and 98\% F1 on average, across various poisoning rates, while the ASR has been reduced to 6\%-21\%.

In this scenario too, the stealthiest version of the attack, using the Combined SHAP strategy, leads to a larger fraction of poisoning points included in $D_{clean}$ before the fixed threshold.
However, as shown in \Cref{fig:lgb_ember_iterative_combined_strat}, \Cref{app:iterative-scoring}, the attack success rate remains relatively low unless high poisoning percentages are used.

\subsubsection{Patching.}

The patching strategy performs relatively poorly at preventing the attack in this scenario, especially at higher poisoning rates. 
Conversely, as in the experiments on network data, it outperforms the pure filtering approach for what concerns preserving model utility. 
In particular, the FPR obtained through patching are consistently better, with an average gain of about 0.5\%, than those obtained with filtering at different poisoning rates, as shown in \Cref{tab:patching80-ember}.

\subsubsection{Loss analysis and sanitization}

\begin{table}[t]

\caption{Comparison of patching and filtering sanitization approaches after loss analysis, using $z_t = 2\sigma$. LightGBM model on the EMBER dataset. Also showing the Base ASR value for the undefended attack. Results are averages of two runs on different random seeds, for different strategies.}
\label{tab:patching-loss-280-ember}

\begin{adjustbox}{width=0.98\linewidth}

\begin{tabular}{lll | c | rrr}

{\bf Sanitization} & \multicolumn{2}{c}{{\bf Attack - Poisoning}} & {\bf Base ASR} & {\bf ASR} & {\bf Test FPR} & {\bf Test F1} \\ \hline

\multirow[c]{8}{*}{Filtering} & \multirow[c]{4}{*}{Independent} & 0.5\% & 24.53\% & 5.60\% & 2.01\% & 98.22\% \\
 &  & 1\% & 34.27\% & 6.51\% & 2.08\% & 98.18\% \\
 &  & 2\% & 59.42\% & 6.38\% & 2.12\% & 98.15\% \\
 &  & 4\% & 98.67\% & 7.03\% & 2.04\% & 98.21\% \\

\cline{2-7}

 & \multirow[c]{4}{*}{Combined} & 0.5\%  & 3.89\% & 1.37\% & 2.08\% & 98.24\% \\
 &  & 1\% & 9.09\% & 1.84\% & 2.29\% & 98.15\% \\
 &  & 2\% & 19.20\% & 1.49\% & 2.22\% & 98.16\% \\
 &  & 4\% & 49.04\% & 1.45\% & 2.43\% & 98.11\% \\

\cline{1-7}

\multirow[c]{8}{*}{Patching} & \multirow[c]{4}{*}{Independent} & 0.5\% & 24.53\% & 19.46\% & 1.68\% & 98.33\% \\
 &  & 1\% & 34.27\% & 36.11\% & 1.68\% & 98.31\% \\
 &  & 2\% & 59.42\% & 47.22\% & 1.67\% & 98.33\% \\
 
\cline{2-7}

 & \multirow[c]{4}{*}{Combined} & 0.5\% & 3.89\% & 19.86\% & 1.65\% & 98.29\% \\
 &  & 1\% & 9.09\% & 36.04\% & 1.76\% & 98.32\% \\
 &  & 2\% & 19.20\% & 36.41\% & 1.84\% & 98.27\% \\

\end{tabular}

\end{adjustbox}

\end{table}

\Cref{tab:patching-loss-280-ember} shows the results of applying our loss analysis to identify the clusters to sanitize.
The results are reported for a threshold Z-score of 2 standard deviations.
This approach reflects the general trends observed before, but improves on the fixed threshold method discussed above when applied with both the filtering and patching sanitization approaches.

\section{Discussion and Limitations}

Defensive mechanisms are always subject to limitations and the general problem of defending from arbitrary poisoning attacks is far from being solved.
With this work we address clean-label backdoor attacks, and propose an effective mitigation approach for existing threats.

Differently from provable defenses, our method is heuristic in nature, trading off protection guarantees for large improvements in retained model utility.
In contrast to other heuristic mitigation approaches introduced in the context of computer vision systems, our method relaxes some strong assumptions on clean data availability and victim model architectures that are difficult to satisfy in the cybersecurity domain.

While we defend against the attack formulations proposed in literature, adaptive attacks against our defense are possible.
For instance, an adversary could change the feature selection process in the attacks to choose a fraction of relevant features and a fraction of non-relevant ones.
While this would likely reduce the attack success rate, it may also reduce the likelihood of poisoning points clustering together in the relevant feature subspace.
Therefore, this would allow the attacker to arbitrarily trade-off attack success to reduce the likelihood of being discovered.

\section{Conclusions}

In this work we introduce a mitigation mechanism against clean-label backdoor poisoning attacks targeted at cybersecurity classifiers.
Our method eliminates many of the most common assumptions used in other defensive techniques against backdoor attacks.
Namely, we remove the need for separate clean datasets, which can be difficult to obtain for cybersecurity tasks, and any assumptions on the architecture of the models used. Our defense effectively reduces the attack success rate in multiple scenarios, while also preserving a high model utility and a low false positives rate.

\section*{Acknowledgements}

This research was sponsored by MIT Lincoln Laboratory,  the NSF awards CNS-2312875 and CNS-2331081, the U.S. Army
Combat Capabilities Development Command Army Research Laboratory (DEVCOM ARL) under Cooperative Agreement Number W911NF-13-2-0045, and the Department of Defense
Multidisciplinary Research Program of the University Research Initiative (MURI) under contract W911NF-21-1-0322. 

\bibliographystyle{ACM-Reference-Format}
\bibliography{prj_defnet, additional, netpois, prj_vpn}


\begin{thebibliography}{60}

%%% the Web address.

\ifx \showCODEN    \undefined \def \showCODEN     #1{\unskip}     \fi
\ifx \showDOI      \undefined \def \showDOI       #1{#1}\fi
\ifx \showISBNx    \undefined \def \showISBNx     #1{\unskip}     \fi
\ifx \showISBNxiii \undefined \def \showISBNxiii  #1{\unskip}     \fi
\ifx \showISSN     \undefined \def \showISSN      #1{\unskip}     \fi
\ifx \showLCCN     \undefined \def \showLCCN      #1{\unskip}     \fi
\ifx \shownote     \undefined \def \shownote      #1{#1}          \fi
\ifx \showarticletitle \undefined \def \showarticletitle #1{#1}   \fi
\ifx \showURL      \undefined \def \showURL       {\relax}        \fi
% invisible to TeX
\providecommand\bibfield[2]{#2}
\providecommand\bibinfo[2]{#2}
\providecommand\natexlab[1]{#1}
\providecommand\showeprint[2][]{arXiv:#2}

\bibitem[Anderson and Roth(2018)]%
        {anderson2018ember}
\bibfield{author}{\bibinfo{person}{Hyrum~S Anderson} {and} \bibinfo{person}{Phil Roth}.} \bibinfo{year}{2018}\natexlab{}.
\newblock \showarticletitle{Ember: an open dataset for training static pe malware machine learning models}.
\newblock \bibinfo{journal}{\emph{arXiv preprint arXiv:1804.04637}} (\bibinfo{year}{2018}).
\newblock


\bibitem[Ankerst et~al\mbox{.}(1999)]%
        {ankerst1999optics}
\bibfield{author}{\bibinfo{person}{Mihael Ankerst}, \bibinfo{person}{Markus~M. Breunig}, \bibinfo{person}{Hans peter Kriegel}, {and} \bibinfo{person}{Jörg Sander}.} \bibinfo{year}{1999}\natexlab{}.
\newblock \showarticletitle{OPTICS: Ordering Points To Identify the Clustering Structure}. \bibinfo{publisher}{ACM Press}, \bibinfo{pages}{49--60}.
\newblock


\bibitem[Antonakakis et~al\mbox{.}(2011)]%
        {antonakakis11}
\bibfield{author}{\bibinfo{person}{Manos Antonakakis}, \bibinfo{person}{Roberto Perdisci}, \bibinfo{person}{Wenke Lee}, \bibinfo{person}{Nikolaos Vasiloglou}, {and} \bibinfo{person}{David Dagon}.} \bibinfo{year}{2011}\natexlab{}.
\newblock \showarticletitle{Detecting Malware Domains at the Upper {DNS} Hierarchy}. In \bibinfo{booktitle}{\emph{Proceedings of the 20th USENIX Conference on Security}} (San Francisco, CA) \emph{(\bibinfo{series}{SEC'11})}. \bibinfo{publisher}{USENIX Association}, \bibinfo{address}{USA}, \bibinfo{pages}{27}.
\newblock


\bibitem[Apruzzese et~al\mbox{.}(2019)]%
        {apruzzeseAddressingAdversarialAttacks2019}
\bibfield{author}{\bibinfo{person}{Giovanni Apruzzese}, \bibinfo{person}{Michele Colajanni}, \bibinfo{person}{Luca Ferretti}, {and} \bibinfo{person}{Mirco Marchetti}.} \bibinfo{year}{2019}\natexlab{}.
\newblock \showarticletitle{Addressing {{Adversarial Attacks Against Security Systems Based}} on {{Machine Learning}}}. In \bibinfo{booktitle}{\emph{2019 11th {{International Conference}} on {{Cyber Conflict}} ({{CyCon}})}}, Vol.~\bibinfo{volume}{900}. \bibinfo{pages}{1--18}.
\newblock
\showISSN{2325-5374}
\urldef\tempurl%
\url{https://doi.org/10.23919/CYCON.2019.8756865}
\showDOI{\tempurl}


\bibitem[ATLAS(1999)]%
        {altasVTPoison}
\bibfield{author}{\bibinfo{person}{MITRE ATLAS}.} \bibinfo{year}{1999}\natexlab{}.
\newblock \bibinfo{booktitle}{\emph{{VirusTotal Poisoning} 2020}}.
\newblock
\urldef\tempurl%
\url{https://atlas.mitre.org/studies/AML.CS0002}
\showURL{%
\tempurl}


\bibitem[Ayub et~al\mbox{.}(2020)]%
        {Ayub2020}
\bibfield{author}{\bibinfo{person}{Md.~Ahsan Ayub}, \bibinfo{person}{William~A. Johnson}, \bibinfo{person}{Douglas~A. Talbert}, {and} \bibinfo{person}{Ambareen Siraj}.} \bibinfo{year}{2020}\natexlab{}.
\newblock \showarticletitle{Model Evasion Attack on Intrusion Detection Systems using Adversarial Machine Learning}. In \bibinfo{booktitle}{\emph{2020 54th Annual Conference on Information Sciences and Systems (CISS)}}. \bibinfo{pages}{1--6}.
\newblock
\urldef\tempurl%
\url{https://doi.org/10.1109/CISS48834.2020.1570617116}
\showDOI{\tempurl}


\bibitem[Biggio et~al\mbox{.}(2012)]%
        {biggio2012poisoning}
\bibfield{author}{\bibinfo{person}{Battista Biggio}, \bibinfo{person}{Blaine Nelson}, {and} \bibinfo{person}{Pavel Laskov}.} \bibinfo{year}{2012}\natexlab{}.
\newblock \showarticletitle{Poisoning attacks against support vector machines}. In \bibinfo{booktitle}{\emph{Proceedings of the 29th International Coference on International Conference on Machine Learning}}. \bibinfo{pages}{1467--1474}.
\newblock


\bibitem[Burkhart et~al\mbox{.}(2010)]%
        {Burkhart2010FeatureStatistics}
\bibfield{author}{\bibinfo{person}{Martin Burkhart}, \bibinfo{person}{Mario Strasser}, \bibinfo{person}{Dilip Many}, {and} \bibinfo{person}{Xenofontas Dimitropoulos}.} \bibinfo{year}{2010}\natexlab{}.
\newblock \showarticletitle{{SEPIA}: {Privacy-Preserving} Aggregation of {Multi-Domain} Network Events and Statistics}. In \bibinfo{booktitle}{\emph{19th USENIX Security Symposium (USENIX Security 10)}}. \bibinfo{publisher}{USENIX Association}, \bibinfo{address}{Washington, DC}.
\newblock
\urldef\tempurl%
\url{https://www.usenix.org/conference/usenixsecurity10/sepia-privacy-preserving-aggregation-multi-domain-network-events-and}
\showURL{%
\tempurl}


\bibitem[Cao and Gong(2017)]%
        {Cao2017}
\bibfield{author}{\bibinfo{person}{Xiaoyu Cao} {and} \bibinfo{person}{Neil~Zhenqiang Gong}.} \bibinfo{year}{2017}\natexlab{}.
\newblock \showarticletitle{Mitigating Evasion Attacks to Deep Neural Networks via Region-Based Classification}. In \bibinfo{booktitle}{\emph{Proceedings of the 33rd Annual Computer Security Applications Conference}} (Orlando, FL, USA) \emph{(\bibinfo{series}{ACSAC '17})}. \bibinfo{publisher}{Association for Computing Machinery}, \bibinfo{address}{New York, NY, USA}, \bibinfo{pages}{278–287}.
\newblock
\showISBNx{9781450353458}
\urldef\tempurl%
\url{https://doi.org/10.1145/3134600.3134606}
\showDOI{\tempurl}


\bibitem[Carlini et~al\mbox{.}(2024)]%
        {carlini2024poisoning}
\bibfield{author}{\bibinfo{person}{Nicholas Carlini}, \bibinfo{person}{Matthew Jagielski}, \bibinfo{person}{Christopher~A Choquette-Choo}, \bibinfo{person}{Daniel Paleka}, \bibinfo{person}{Will Pearce}, \bibinfo{person}{Hyrum Anderson}, \bibinfo{person}{Andreas Terzis}, \bibinfo{person}{Kurt Thomas}, {and} \bibinfo{person}{Florian Tram{\`e}r}.} \bibinfo{year}{2024}\natexlab{}.
\newblock \showarticletitle{Poisoning Web-Scale Training Datasets is Practical}. In \bibinfo{booktitle}{\emph{2024 IEEE Symposium on Security and Privacy (SP)}}. IEEE Computer Society, \bibinfo{pages}{176--176}.
\newblock


\bibitem[Chen et~al\mbox{.}(2018)]%
        {chen2018detecting}
\bibfield{author}{\bibinfo{person}{Bryant Chen}, \bibinfo{person}{Wilka Carvalho}, \bibinfo{person}{Nathalie Baracaldo}, \bibinfo{person}{Heiko Ludwig}, \bibinfo{person}{Benjamin Edwards}, \bibinfo{person}{Taesung Lee}, \bibinfo{person}{Ian Molloy}, {and} \bibinfo{person}{Biplav Srivastava}.} \bibinfo{year}{2018}\natexlab{}.
\newblock \bibinfo{title}{Detecting Backdoor Attacks on Deep Neural Networks by Activation Clustering}.
\newblock
\newblock
\showeprint[arxiv]{1811.03728}~[cs.LG]


\bibitem[Chernikova and Oprea(2022)]%
        {FENCE}
\bibfield{author}{\bibinfo{person}{Alesia Chernikova} {and} \bibinfo{person}{Alina Oprea}.} \bibinfo{year}{2022}\natexlab{}.
\newblock \showarticletitle{FENCE: Feasible Evasion Attacks on Neural Networks in Constrained Environments}.
\newblock \bibinfo{journal}{\emph{ACM Trans. Priv. Secur.}} \bibinfo{volume}{25}, \bibinfo{number}{4}, Article \bibinfo{articleno}{34} (\bibinfo{date}{jul} \bibinfo{year}{2022}), \bibinfo{numpages}{34}~pages.
\newblock
\showISSN{2471-2566}
\urldef\tempurl%
\url{https://doi.org/10.1145/3544746}
\showDOI{\tempurl}


\bibitem[Ester et~al\mbox{.}(1996)]%
        {dbscan1996}
\bibfield{author}{\bibinfo{person}{Martin Ester}, \bibinfo{person}{Hans-Peter Kriegel}, \bibinfo{person}{J\"{o}rg Sander}, {and} \bibinfo{person}{Xiaowei Xu}.} \bibinfo{year}{1996}\natexlab{}.
\newblock \showarticletitle{A density-based algorithm for discovering clusters in large spatial databases with noise}. In \bibinfo{booktitle}{\emph{Proceedings of the Second International Conference on Knowledge Discovery and Data Mining}} (Portland, Oregon) \emph{(\bibinfo{series}{KDD'96})}. \bibinfo{publisher}{AAAI Press}, \bibinfo{pages}{226–231}.
\newblock


\bibitem[Garc{\'i}a et~al\mbox{.}(2014)]%
        {garciaEmpiricalComparisonBotnet2014}
\bibfield{author}{\bibinfo{person}{S. Garc{\'i}a}, \bibinfo{person}{M. Grill}, \bibinfo{person}{J. Stiborek}, {and} \bibinfo{person}{A. Zunino}.} \bibinfo{year}{2014}\natexlab{}.
\newblock \showarticletitle{An Empirical Comparison of Botnet Detection Methods}.
\newblock \bibinfo{journal}{\emph{Computers and Security}}  \bibinfo{volume}{45} (\bibinfo{date}{Sept.} \bibinfo{year}{2014}), \bibinfo{pages}{100--123}.
\newblock
\showISSN{0167-4048}
\urldef\tempurl%
\url{https://doi.org/10.1016/j.cose.2014.05.011}
\showDOI{\tempurl}


\bibitem[Gu et~al\mbox{.}(2019)]%
        {guBadNetsEvaluatingBackdooring2019}
\bibfield{author}{\bibinfo{person}{T. Gu}, \bibinfo{person}{K. Liu}, \bibinfo{person}{B. {Dolan-Gavitt}}, {and} \bibinfo{person}{S. Garg}.} \bibinfo{year}{2019}\natexlab{}.
\newblock \showarticletitle{{{BadNets}}: {{Evaluating Backdooring Attacks}} on {{Deep Neural Networks}}}.
\newblock \bibinfo{journal}{\emph{IEEE Access}}  \bibinfo{volume}{7} (\bibinfo{year}{2019}), \bibinfo{pages}{47230--47244}.
\newblock
\showISSN{2169-3536}
\urldef\tempurl%
\url{https://doi.org/10.1109/ACCESS.2019.2909068}
\showDOI{\tempurl}


\bibitem[Handley et~al\mbox{.}(2001)]%
        {Handley2001}
\bibfield{author}{\bibinfo{person}{Mark Handley}, \bibinfo{person}{Vern Paxson}, {and} \bibinfo{person}{Christian Kreibich}.} \bibinfo{year}{2001}\natexlab{}.
\newblock \showarticletitle{Network Intrusion Detection: Evasion, Traffic Normalization, and {End-to-End} Protocol Semantics}. In \bibinfo{booktitle}{\emph{10th USENIX Security Symposium (USENIX Security 01)}}. \bibinfo{publisher}{USENIX Association}, \bibinfo{address}{Washington, D.C.}
\newblock
\urldef\tempurl%
\url{https://www.usenix.org/conference/10th-usenix-security-symposium/network-intrusion-detection-evasion-traffic-normalization}
\showURL{%
\tempurl}


\bibitem[Hayase et~al\mbox{.}(2021)]%
        {Spectre}
\bibfield{author}{\bibinfo{person}{Jonathan Hayase}, \bibinfo{person}{Weihao Kong}, \bibinfo{person}{Raghav Somani}, {and} \bibinfo{person}{Sewoong Oh}.} \bibinfo{year}{2021}\natexlab{}.
\newblock \showarticletitle{SPECTRE: defending against backdoor attacks using robust statistics}. In \bibinfo{booktitle}{\emph{Proceedings of the 38th International Conference on Machine Learning}} \emph{(\bibinfo{series}{Proceedings of Machine Learning Research}, Vol.~\bibinfo{volume}{139})}, \bibfield{editor}{\bibinfo{person}{Marina Meila} {and} \bibinfo{person}{Tong Zhang}} (Eds.). \bibinfo{publisher}{PMLR}, \bibinfo{pages}{4129--4139}.
\newblock
\urldef\tempurl%
\url{https://proceedings.mlr.press/v139/hayase21a.html}
\showURL{%
\tempurl}


\bibitem[Heng and Soh(2023)]%
        {Heng2023SEAM}
\bibfield{author}{\bibinfo{person}{Alvin Heng} {and} \bibinfo{person}{Harold Soh}.} \bibinfo{year}{2023}\natexlab{}.
\newblock \showarticletitle{Selective Amnesia: A Continual Learning Approach to Forgetting in Deep Generative Models}. In \bibinfo{booktitle}{\emph{NeurIPS}}, Vol.~\bibinfo{volume}{36}. \bibinfo{pages}{17170--17194}.
\newblock


\bibitem[Ho et~al\mbox{.}(2022)]%
        {NestedTraining_Clean}
\bibfield{author}{\bibinfo{person}{Samson Ho}, \bibinfo{person}{Achyut Reddy}, \bibinfo{person}{Sridhar Venkatesan}, \bibinfo{person}{Rauf Izmailov}, \bibinfo{person}{Ritu Chadha}, {and} \bibinfo{person}{Alina Oprea}.} \bibinfo{year}{2022}\natexlab{}.
\newblock \showarticletitle{Data Sanitization Approach to Mitigate Clean-Label Attacks Against Malware Detection Systems}. In \bibinfo{booktitle}{\emph{MILCOM 2022 - 2022 IEEE Military Communications Conference (MILCOM)}}. \bibinfo{pages}{993--998}.
\newblock
\urldef\tempurl%
\url{https://doi.org/10.1109/MILCOM55135.2022.10017768}
\showDOI{\tempurl}


\bibitem[Holodnak et~al\mbox{.}(2022)]%
        {holodnakBackdoorPoisoningEncrypted2022}
\bibfield{author}{\bibinfo{person}{John~T. Holodnak}, \bibinfo{person}{Olivia Brown}, \bibinfo{person}{Jason Matterer}, {and} \bibinfo{person}{Andrew Lemke}.} \bibinfo{year}{2022}\natexlab{}.
\newblock \showarticletitle{Backdoor {{Poisoning}} of {{Encrypted Traffic Classifiers}}}. In \bibinfo{booktitle}{\emph{2022 {{IEEE International Conference}} on {{Data Mining Workshops}} ({{ICDMW}})}}. \bibinfo{pages}{577--585}.
\newblock
\showISSN{2375-9259}
\urldef\tempurl%
\url{https://doi.org/10.1109/ICDMW58026.2022.00080}
\showDOI{\tempurl}


\bibitem[IBM(2023)]%
        {ibmqradar}
\bibfield{author}{\bibinfo{person}{IBM}.} \bibinfo{year}{2023}\natexlab{}.
\newblock \bibinfo{title}{{IBM Security QRadar XDR}}.
\newblock
\newblock
\newblock
\shownote{{https://www.ibm.com/qradar}}.


\bibitem[Ingalls(2021)]%
        {xdr_solutions}
\bibfield{author}{\bibinfo{person}{Sam Ingalls}.} \bibinfo{year}{2021}\natexlab{}.
\newblock \bibinfo{title}{{Top XDR Security Solutions for 2022}}.
\newblock \bibinfo{howpublished}{\url{https://www.esecurityplanet.com/products/xdr-security-solutions/}}.
\newblock


\bibitem[Invernizzi et~al\mbox{.}(2014)]%
        {nazca_ndss2014}
\bibfield{author}{\bibinfo{person}{Luca Invernizzi}, \bibinfo{person}{Sung ju Lee}, \bibinfo{person}{Stanislav Miskovic}, \bibinfo{person}{Marco Mellia}, \bibinfo{person}{Ruben Torres}, \bibinfo{person}{Christopher Kruegel}, \bibinfo{person}{Sabyasachi Saha}, {and} \bibinfo{person}{Giovanni Vigna}.} \bibinfo{year}{2014}\natexlab{}.
\newblock \showarticletitle{{Nazca: Detecting Malware Distribution in Large-Scale Networks}}. In \bibinfo{booktitle}{\emph{NDSS}}.
\newblock


\bibitem[Levine and Feizi(2021)]%
        {DPA2021}
\bibfield{author}{\bibinfo{person}{Alexander Levine} {and} \bibinfo{person}{Soheil Feizi}.} \bibinfo{year}{2021}\natexlab{}.
\newblock \showarticletitle{Deep Partition Aggregation: Provable Defenses against General Poisoning Attacks}. In \bibinfo{booktitle}{\emph{9th International Conference on Learning Representations, {ICLR} 2021, Virtual Event, Austria, May 3-7, 2021}}. \bibinfo{publisher}{OpenReview.net}.
\newblock
\urldef\tempurl%
\url{https://openreview.net/forum?id=YUGG2tFuPM}
\showURL{%
\tempurl}


\bibitem[Li et~al\mbox{.}(2018)]%
        {Li2018EPD}
\bibfield{author}{\bibinfo{person}{Pan Li}, \bibinfo{person}{Qiang Liu}, \bibinfo{person}{Wentao Zhao}, \bibinfo{person}{Dongxu Wang}, {and} \bibinfo{person}{Siqi Wang}.} \bibinfo{year}{2018}\natexlab{}.
\newblock \showarticletitle{Chronic Poisoning against Machine Learning Based IDSs Using Edge Pattern Detection}. In \bibinfo{booktitle}{\emph{2018 IEEE International Conference on Communications (ICC)}}. \bibinfo{pages}{1--7}.
\newblock
\urldef\tempurl%
\url{https://doi.org/10.1109/ICC.2018.8422328}
\showDOI{\tempurl}


\bibitem[Li et~al\mbox{.}(2021)]%
        {Li2021NeuralAt}
\bibfield{author}{\bibinfo{person}{Yige Li}, \bibinfo{person}{Xixiang Lyu}, \bibinfo{person}{Nodens Koren}, \bibinfo{person}{Lingjuan Lyu}, \bibinfo{person}{Bo Li}, {and} \bibinfo{person}{Xingjun Ma}.} \bibinfo{year}{2021}\natexlab{}.
\newblock \showarticletitle{Neural Attention Distillation: Erasing Backdoor Triggers from Deep Neural Networks}. In \bibinfo{booktitle}{\emph{9th International Conference on Learning Representations, {ICLR} 2021, Virtual Event, Austria, May 3-7, 2021}}. \bibinfo{publisher}{OpenReview.net}.
\newblock


\bibitem[Liu et~al\mbox{.}(2018)]%
        {liu2018fine-pruning}
\bibfield{author}{\bibinfo{person}{Kang Liu}, \bibinfo{person}{Brendan Dolan-Gavitt}, {and} \bibinfo{person}{Siddharth Garg}.} \bibinfo{year}{2018}\natexlab{}.
\newblock \showarticletitle{Fine-Pruning: Defending Against Backdooring Attacks on Deep Neural Networks}. In \bibinfo{booktitle}{\emph{Research in Attacks, Intrusions, and Defenses}}. \bibinfo{pages}{273--294}.
\newblock


\bibitem[Liu et~al\mbox{.}(2019)]%
        {ABS}
\bibfield{author}{\bibinfo{person}{Yingqi Liu}, \bibinfo{person}{Wen-Chuan Lee}, \bibinfo{person}{Guanhong Tao}, \bibinfo{person}{Shiqing Ma}, \bibinfo{person}{Yousra Aafer}, {and} \bibinfo{person}{Xiangyu Zhang}.} \bibinfo{year}{2019}\natexlab{}.
\newblock \showarticletitle{{ABS}: Scanning Neural Networks for Back-doors by Artificial Brain Stimulation}. In \bibinfo{booktitle}{\emph{Proceedings of the 2019 ACM SIGSAC Conference on Computer and Communications Security}} (London, United Kingdom) \emph{(\bibinfo{series}{CCS '19})}. \bibinfo{publisher}{Association for Computing Machinery}, \bibinfo{address}{New York, NY, USA}, \bibinfo{pages}{1265–1282}.
\newblock
\showISBNx{9781450367479}
\urldef\tempurl%
\url{https://doi.org/10.1145/3319535.3363216}
\showDOI{\tempurl}


\bibitem[Lundberg and Lee(2017)]%
        {SHAP}
\bibfield{author}{\bibinfo{person}{Scott~M Lundberg} {and} \bibinfo{person}{Su-In Lee}.} \bibinfo{year}{2017}\natexlab{}.
\newblock \showarticletitle{A Unified Approach to Interpreting Model Predictions}. In \bibinfo{booktitle}{\emph{Advances in Neural Information Processing Systems}}, \bibfield{editor}{\bibinfo{person}{I.~Guyon}, \bibinfo{person}{U.~Von Luxburg}, \bibinfo{person}{S.~Bengio}, \bibinfo{person}{H.~Wallach}, \bibinfo{person}{R.~Fergus}, \bibinfo{person}{S.~Vishwanathan}, {and} \bibinfo{person}{R.~Garnett}} (Eds.), Vol.~\bibinfo{volume}{30}. \bibinfo{publisher}{Curran Associates, Inc.}
\newblock
\urldef\tempurl%
\url{https://proceedings.neurips.cc/paper_files/paper/2017/file/8a20a8621978632d76c43dfd28b67767-Paper.pdf}
\showURL{%
\tempurl}


\bibitem[MalwareGuard FireEye(2020)]%
        {fireeye2020Jun}
MalwareGuard FireEye \bibinfo{year}{2020}\natexlab{}.
\newblock \bibinfo{title}{{MalwareGuard: FireEye{'}s Machine Learning Model to Detect and Prevent Malware}}.
\newblock
\newblock
\newblock
\shownote{{https://www.fireeye.com/blog/products-and-services/2018/07/malwareguard-fireeye-machine-learning-model-to-detect-and-prevent-malware.html}}.


\bibitem[Microsoft(2021)]%
        {microsoftdefender}
\bibfield{author}{\bibinfo{person}{Microsoft}.} \bibinfo{year}{2021}\natexlab{}.
\newblock \bibinfo{title}{{Microsoft Defender for Endpoint {$\vert$} Microsoft Security}}.
\newblock
\newblock
\urldef\tempurl%
\url{https://www.microsoft.com/en-us/security/business/threat-protection/endpoint-defender}
\showURL{%
\tempurl}


\bibitem[Mirsky et~al\mbox{.}(2018)]%
        {mirskyKitsuneEnsembleAutoencoders2018a}
\bibfield{author}{\bibinfo{person}{Yisroel Mirsky}, \bibinfo{person}{Tomer Doitshman}, \bibinfo{person}{Yuval Elovici}, {and} \bibinfo{person}{Asaf Shabtai}.} \bibinfo{year}{2018}\natexlab{}.
\newblock \showarticletitle{Kitsune: {{An Ensemble}} of {{Autoencoders}} for {{Online Network Intrusion Detection}}}. In \bibinfo{booktitle}{\emph{Proceedings 2018 {{Network}} and {{Distributed System Security Symposium}}}}. \bibinfo{publisher}{{Internet Society}}, \bibinfo{address}{{San Diego, CA}}.
\newblock
\showISBNx{978-1-891562-49-5}
\urldef\tempurl%
\url{https://doi.org/10.14722/ndss.2018.23204}
\showDOI{\tempurl}


\bibitem[Mo et~al\mbox{.}(2024)]%
        {TED}
\bibfield{author}{\bibinfo{person}{X. Mo}, \bibinfo{person}{Y. Zhang}, \bibinfo{person}{L. Zhang}, \bibinfo{person}{W. Luo}, \bibinfo{person}{N. Sun}, \bibinfo{person}{S. Hu}, \bibinfo{person}{S. Gao}, {and} \bibinfo{person}{Y. Xiang}.} \bibinfo{year}{2024}\natexlab{}.
\newblock \showarticletitle{Robust Backdoor Detection for Deep Learning via Topological Evolution Dynamics}. In \bibinfo{booktitle}{\emph{2024 IEEE Symposium on Security and Privacy (SP)}}. \bibinfo{publisher}{IEEE Computer Society}, \bibinfo{address}{Los Alamitos, CA, USA}, \bibinfo{pages}{174--174}.
\newblock
\showISSN{2375-1207}
\urldef\tempurl%
\url{https://doi.org/10.1109/SP54263.2024.00174}
\showDOI{\tempurl}


\bibitem[Moore et~al\mbox{.}(2005)]%
        {mooreDiscriminatorsUseFlowbased2005}
\bibfield{author}{\bibinfo{person}{Andrew Moore}, \bibinfo{person}{Denis Zuev}, {and} \bibinfo{person}{Michael Crogan}.} \bibinfo{year}{2005}\natexlab{}.
\newblock \bibinfo{booktitle}{\emph{Discriminators for Use in Flow-Based Classification}}.
\newblock \bibinfo{type}{{T}echnical {R}eport}. \bibinfo{institution}{{Queen Mary and Westfield College, Department of Computer Science}}.
\newblock


\bibitem[Ning et~al\mbox{.}(2022)]%
        {ningTrojanFlowNeuralBackdoor2022}
\bibfield{author}{\bibinfo{person}{Rui Ning}, \bibinfo{person}{Chunsheng Xin}, {and} \bibinfo{person}{Hongyi Wu}.} \bibinfo{year}{2022}\natexlab{}.
\newblock \showarticletitle{{{TrojanFlow}}: {{A Neural Backdoor Attack}} to {{Deep Learning-based Network Traffic Classifiers}}}. In \bibinfo{booktitle}{\emph{{{IEEE INFOCOM}} 2022 - {{IEEE Conference}} on {{Computer Communications}}}}. \bibinfo{pages}{1429--1438}.
\newblock
\showISSN{2641-9874}
\urldef\tempurl%
\url{https://doi.org/10.1109/INFOCOM48880.2022.9796878}
\showDOI{\tempurl}


\bibitem[Ongun et~al\mbox{.}(2021)]%
        {ongunPORTFILERPortLevelNetwork2021}
\bibfield{author}{\bibinfo{person}{Talha Ongun}, \bibinfo{person}{Oliver Spohngellert}, \bibinfo{person}{Benjamin Miller}, \bibinfo{person}{Simona Boboila}, \bibinfo{person}{Alina Oprea}, \bibinfo{person}{Tina {Eliassi-Rad}}, \bibinfo{person}{Jason Hiser}, \bibinfo{person}{Alastair Nottingham}, \bibinfo{person}{Jack Davidson}, {and} \bibinfo{person}{Malathi Veeraraghavan}.} \bibinfo{year}{2021}\natexlab{}.
\newblock \showarticletitle{{{PORTFILER}}: {{Port-Level Network Profiling}} for {{Self-Propagating Malware Detection}}}. In \bibinfo{booktitle}{\emph{2021 {{IEEE Conference}} on {{Communications}} and {{Network Security}} ({{CNS}})}}. \bibinfo{pages}{182--190}.
\newblock
\urldef\tempurl%
\url{https://doi.org/10.1109/CNS53000.2021.9705045}
\showDOI{\tempurl}


\bibitem[Oprea et~al\mbox{.}(2018)]%
        {MADE}
\bibfield{author}{\bibinfo{person}{Alina Oprea}, \bibinfo{person}{Zhou Li}, \bibinfo{person}{Robin Norris}, {and} \bibinfo{person}{Kevin Bowers}.} \bibinfo{year}{2018}\natexlab{}.
\newblock \showarticletitle{{MADE}: Security Analytics for Enterprise Threat Detection}. In \bibinfo{booktitle}{\emph{Proceedings of Annual Computer Security Applications Conference}} \emph{(\bibinfo{series}{ACSAC})}.
\newblock
\urldef\tempurl%
\url{https://doi.org/10.1145/3274694.3274710}
\showDOI{\tempurl}


\bibitem[Oprea and Vassilev(2023)]%
        {opreaAdversarialMachineLearning2023}
\bibfield{author}{\bibinfo{person}{Alina Oprea} {and} \bibinfo{person}{Apostol Vassilev}.} \bibinfo{year}{2023}\natexlab{}.
\newblock \bibinfo{booktitle}{\emph{Adversarial {{Machine Learning}}: {{A Taxonomy}} and {{Terminology}} of {{Attacks}} and {{Mitigations}} ({{Draft}})}}.
\newblock \bibinfo{type}{{T}echnical {R}eport} NIST AI 100-2e2023 ipd. \bibinfo{institution}{{National Institute of Standards and Technology}}.
\newblock


\bibitem[Papadopoulos et~al\mbox{.}(2021a)]%
        {papadopoulosLaunchingAdversarialAttacks2021}
\bibfield{author}{\bibinfo{person}{Pavlos Papadopoulos}, \bibinfo{person}{Oliver {Thornewill von Essen}}, \bibinfo{person}{Nikolaos Pitropakis}, \bibinfo{person}{Christos Chrysoulas}, \bibinfo{person}{Alexios Mylonas}, {and} \bibinfo{person}{William~J. Buchanan}.} \bibinfo{year}{2021}\natexlab{a}.
\newblock \showarticletitle{Launching {{Adversarial Attacks}} against {{Network Intrusion Detection Systems}} for {{IoT}}}.
\newblock \bibinfo{journal}{\emph{Journal of Cybersecurity and Privacy}} \bibinfo{volume}{1}, \bibinfo{number}{2} (\bibinfo{date}{June} \bibinfo{year}{2021}), \bibinfo{pages}{252--273}.
\newblock
\showISSN{2624-800X}
\urldef\tempurl%
\url{https://doi.org/10.3390/jcp1020014}
\showDOI{\tempurl}


\bibitem[Papadopoulos et~al\mbox{.}(2021b)]%
        {Papadopoulos2021}
\bibfield{author}{\bibinfo{person}{Pavlos Papadopoulos}, \bibinfo{person}{Oliver Thornewill~von Essen}, \bibinfo{person}{Nikolaos Pitropakis}, \bibinfo{person}{Christos Chrysoulas}, \bibinfo{person}{Alexios Mylonas}, {and} \bibinfo{person}{William~J. Buchanan}.} \bibinfo{year}{2021}\natexlab{b}.
\newblock \showarticletitle{{Launching Adversarial Attacks against Network Intrusion Detection Systems for IoT}}.
\newblock \bibinfo{journal}{\emph{Journal of Cybersecurity and Privacy}} \bibinfo{volume}{1}, \bibinfo{number}{2} (\bibinfo{year}{2021}), \bibinfo{pages}{252--273}.
\newblock
\showISSN{2624-800X}
\urldef\tempurl%
\url{https://doi.org/10.3390/jcp1020014}
\showDOI{\tempurl}


\bibitem[Paxson(1999)]%
        {zeek}
\bibfield{author}{\bibinfo{person}{Vern Paxson}.} \bibinfo{year}{1999}\natexlab{}.
\newblock \showarticletitle{{Bro: a System for Detecting Network Intruders in Real-Time}}.
\newblock \bibinfo{journal}{\emph{Computer Networks}} \bibinfo{volume}{31}, \bibinfo{number}{23-24} (\bibinfo{year}{1999}), \bibinfo{pages}{2435--2463}.
\newblock
\urldef\tempurl%
\url{http://www.icir.org/vern/papers/bro-CN99.pdf}
\showURL{%
\tempurl}


\bibitem[Qi et~al\mbox{.}(2023)]%
        {qiProactiveMLApproach2023}
\bibfield{author}{\bibinfo{person}{Xiangyu Qi}, \bibinfo{person}{Tinghao Xie}, \bibinfo{person}{Jiachen~T. Wang}, \bibinfo{person}{Tong Wu}, \bibinfo{person}{Saeed Mahloujifar}, {and} \bibinfo{person}{Prateek Mittal}.} \bibinfo{year}{2023}\natexlab{}.
\newblock \showarticletitle{Towards {{A Proactive}} \{\vphantom\}{{ML}}\vphantom\{\} {{Approach}} for {{Detecting Backdoor Poison Samples}}}. In \bibinfo{booktitle}{\emph{32nd {{USENIX Security Symposium}} ({{USENIX Security}} 23)}}. \bibinfo{pages}{1685--1702}.
\newblock
\showISBNx{978-1-939133-37-3}


\bibitem[Severi et~al\mbox{.}(2023)]%
        {severiPoisoningNetworkFlow2023}
\bibfield{author}{\bibinfo{person}{Giorgio Severi}, \bibinfo{person}{Simona Boboila}, \bibinfo{person}{Alina Oprea}, \bibinfo{person}{John Holodnak}, \bibinfo{person}{Kendra Kratkiewicz}, {and} \bibinfo{person}{Jason Matterer}.} \bibinfo{year}{2023}\natexlab{}.
\newblock \showarticletitle{Poisoning {{Network Flow Classifiers}}}. In \bibinfo{booktitle}{\emph{Proceedings of the 39th {{Annual Computer Security Applications Conference}}}} \emph{(\bibinfo{series}{{{ACSAC}} '23})}. \bibinfo{publisher}{Association for Computing Machinery}, \bibinfo{address}{New York, NY, USA}, \bibinfo{pages}{337--351}.
\newblock
\showISBNx{9798400708862}
\urldef\tempurl%
\url{https://doi.org/10.1145/3627106.3627123}
\showDOI{\tempurl}


\bibitem[Severi et~al\mbox{.}(2021)]%
        {severiExplanationGuidedBackdoorPoisoning2021}
\bibfield{author}{\bibinfo{person}{Giorgio Severi}, \bibinfo{person}{Jim Meyer}, \bibinfo{person}{Scott Coull}, {and} \bibinfo{person}{Alina Oprea}.} \bibinfo{year}{2021}\natexlab{}.
\newblock \showarticletitle{Explanation-{{Guided Backdoor Poisoning Attacks Against Malware Classifiers}}}. In \bibinfo{booktitle}{\emph{30th {{USENIX Security Symposium}} ({{USENIX Security}} 21)}}. \bibinfo{pages}{1487--1504}.
\newblock
\showISBNx{978-1-939133-24-3}


\bibitem[Shafahi et~al\mbox{.}(2018)]%
        {shafahiPoisonFrogsTargeted2018}
\bibfield{author}{\bibinfo{person}{Ali Shafahi}, \bibinfo{person}{W.~Ronny Huang}, \bibinfo{person}{Mahyar Najibi}, \bibinfo{person}{Octavian Suciu}, \bibinfo{person}{Christoph Studer}, \bibinfo{person}{Tudor Dumitras}, {and} \bibinfo{person}{Tom Goldstein}.} \bibinfo{year}{2018}\natexlab{}.
\newblock \showarticletitle{Poison {{Frogs}}! {{Targeted Clean-Label Poisoning Attacks}} on {{Neural Networks}}}. In \bibinfo{booktitle}{\emph{Advances in {{Neural Information Processing Systems}}}}.
\newblock


\bibitem[Siva~Kumar et~al\mbox{.}(2020)]%
        {sivakumarAdversarialMachineLearningIndustry2020}
\bibfield{author}{\bibinfo{person}{Ram~Shankar Siva~Kumar}, \bibinfo{person}{Magnus Nystr{\"o}m}, \bibinfo{person}{John Lambert}, \bibinfo{person}{Andrew Marshall}, \bibinfo{person}{Mario Goertzel}, \bibinfo{person}{Andi Comissoneru}, \bibinfo{person}{Matt Swann}, {and} \bibinfo{person}{Sharon Xia}.} \bibinfo{year}{2020}\natexlab{}.
\newblock \showarticletitle{Adversarial {{Machine Learning-Industry Perspectives}}}. In \bibinfo{booktitle}{\emph{2020 {{IEEE Security}} and {{Privacy Workshops}} ({{SPW}})}}. \bibinfo{pages}{69--75}.
\newblock
\urldef\tempurl%
\url{https://doi.org/10.1109/SPW50608.2020.00028}
\showDOI{\tempurl}


\bibitem[Steinbach et~al\mbox{.}(2004)]%
        {Steinbach2004}
\bibfield{author}{\bibinfo{person}{Michael Steinbach}, \bibinfo{person}{Levent Ert{\"o}z}, {and} \bibinfo{person}{Vipin Kumar}.} \bibinfo{year}{2004}\natexlab{}.
\newblock \bibinfo{booktitle}{\emph{The Challenges of Clustering High Dimensional Data}}.
\newblock \bibinfo{publisher}{Springer Berlin Heidelberg}, \bibinfo{address}{Berlin, Heidelberg}, \bibinfo{pages}{273--309}.
\newblock
\showISBNx{978-3-662-08968-2}
\urldef\tempurl%
\url{https://doi.org/10.1007/978-3-662-08968-2_16}
\showDOI{\tempurl}


\bibitem[Steinhardt et~al\mbox{.}(2017)]%
        {Steinhardt2017Cert}
\bibfield{author}{\bibinfo{person}{Jacob Steinhardt}, \bibinfo{person}{Pang~Wei Koh}, {and} \bibinfo{person}{Percy Liang}.} \bibinfo{year}{2017}\natexlab{}.
\newblock \showarticletitle{Certified defenses for data poisoning attacks}. In \bibinfo{booktitle}{\emph{Proceedings of the 31st International Conference on Neural Information Processing Systems}} (Long Beach, California, USA) \emph{(\bibinfo{series}{NIPS'17})}. \bibinfo{publisher}{Curran Associates Inc.}, \bibinfo{address}{Red Hook, NY, USA}, \bibinfo{pages}{3520–3532}.
\newblock
\showISBNx{9781510860964}


\bibitem[Szegedy et~al\mbox{.}(2014)]%
        {DBLP:journals/corr/SzegedyZSBEGF13}
\bibfield{author}{\bibinfo{person}{Christian Szegedy}, \bibinfo{person}{Wojciech Zaremba}, \bibinfo{person}{Ilya Sutskever}, \bibinfo{person}{Joan Bruna}, \bibinfo{person}{Dumitru Erhan}, \bibinfo{person}{Ian~J. Goodfellow}, {and} \bibinfo{person}{Rob Fergus}.} \bibinfo{year}{2014}\natexlab{}.
\newblock \showarticletitle{Intriguing properties of neural networks}. In \bibinfo{booktitle}{\emph{2nd International Conference on Learning Representations, {ICLR} 2014, Banff, AB, Canada, April 14-16, 2014, Conference Track Proceedings}}, \bibfield{editor}{\bibinfo{person}{Yoshua Bengio} {and} \bibinfo{person}{Yann LeCun}} (Eds.).
\newblock
\urldef\tempurl%
\url{http://arxiv.org/abs/1312.6199}
\showURL{%
\tempurl}


\bibitem[Tamersoy et~al\mbox{.}(2014)]%
        {tamersoy14}
\bibfield{author}{\bibinfo{person}{Acar Tamersoy}, \bibinfo{person}{Kevin Roundy}, {and} \bibinfo{person}{Duen~Horng Chau}.} \bibinfo{year}{2014}\natexlab{}.
\newblock \showarticletitle{Guilt by Association: Large Scale Malware Detection by Mining File-Relation Graphs}. In \bibinfo{booktitle}{\emph{Proceedings of the 20th ACM SIGKDD International Conference on Knowledge Discovery and Data Mining}} (New York, New York, USA) \emph{(\bibinfo{series}{KDD '14})}. \bibinfo{publisher}{Association for Computing Machinery}, \bibinfo{address}{New York, NY, USA}, \bibinfo{pages}{1524–1533}.
\newblock
\showISBNx{9781450329569}
\urldef\tempurl%
\url{https://doi.org/10.1145/2623330.2623342}
\showDOI{\tempurl}


\bibitem[Tran et~al\mbox{.}(2018)]%
        {Tran2018SpectralSI}
\bibfield{author}{\bibinfo{person}{Brandon Tran}, \bibinfo{person}{Jerry Li}, {and} \bibinfo{person}{Aleksander Madry}.} \bibinfo{year}{2018}\natexlab{}.
\newblock \showarticletitle{Spectral Signatures in Backdoor Attacks}. In \bibinfo{booktitle}{\emph{Neural Information Processing Systems}}.
\newblock
\urldef\tempurl%
\url{https://api.semanticscholar.org/CorpusID:53298804}
\showURL{%
\tempurl}


\bibitem[Turner et~al\mbox{.}(2019)]%
        {turner2019labelconsistent}
\bibfield{author}{\bibinfo{person}{Alexander Turner}, \bibinfo{person}{Dimitris Tsipras}, {and} \bibinfo{person}{Aleksander Madry}.} \bibinfo{year}{2019}\natexlab{}.
\newblock \bibinfo{title}{Label-Consistent Backdoor Attacks}.
\newblock
\newblock
\showeprint[arxiv]{1912.02771}~[stat.ML]


\bibitem[Venkatesan et~al\mbox{.}(2021)]%
        {NestedTraining}
\bibfield{author}{\bibinfo{person}{Sridhar Venkatesan}, \bibinfo{person}{Harshvardhan Sikka}, \bibinfo{person}{Rauf Izmailov}, \bibinfo{person}{Ritu Chadha}, \bibinfo{person}{Alina Oprea}, {and} \bibinfo{person}{Michael~J. de Lucia}.} \bibinfo{year}{2021}\natexlab{}.
\newblock \showarticletitle{Poisoning Attacks and Data Sanitization Mitigations for Machine Learning Models in Network Intrusion Detection Systems}. In \bibinfo{booktitle}{\emph{MILCOM 2021 - 2021 IEEE Military Communications Conference (MILCOM)}}. \bibinfo{pages}{874--879}.
\newblock
\urldef\tempurl%
\url{https://doi.org/10.1109/MILCOM52596.2021.9652916}
\showDOI{\tempurl}


\bibitem[Wang et~al\mbox{.}(2020)]%
        {Wang2020}
\bibfield{author}{\bibinfo{person}{B. Wang}, \bibinfo{person}{X. Cao}, \bibinfo{person}{J. Jia}, {and} \bibinfo{person}{N.~Z. Gong}.} \bibinfo{year}{2020}\natexlab{}.
\newblock \showarticletitle{n certifying robustness against backdoor attacks via randomized smoothing}. In \bibinfo{booktitle}{\emph{CVPR 2020 Workshop on Adversarial Machine Learning in Computer Vision}}.
\newblock


\bibitem[Wang et~al\mbox{.}(2019)]%
        {NeuralCleanse}
\bibfield{author}{\bibinfo{person}{Bolun Wang}, \bibinfo{person}{Yuanshun Yao}, \bibinfo{person}{Shawn Shan}, \bibinfo{person}{Huiying Li}, \bibinfo{person}{Bimal Viswanath}, \bibinfo{person}{Haitao Zheng}, {and} \bibinfo{person}{Ben~Y. Zhao}.} \bibinfo{year}{2019}\natexlab{}.
\newblock \showarticletitle{{Neural Cleanse}: Identifying and Mitigating Backdoor Attacks in Neural Networks}. In \bibinfo{booktitle}{\emph{2019 IEEE Symposium on Security and Privacy (SP)}}. \bibinfo{pages}{707--723}.
\newblock
\urldef\tempurl%
\url{https://doi.org/10.1109/SP.2019.00031}
\showDOI{\tempurl}


\bibitem[Wang et~al\mbox{.}(2022)]%
        {DFA2022}
\bibfield{author}{\bibinfo{person}{Wenxiao Wang}, \bibinfo{person}{Alexander Levine}, {and} \bibinfo{person}{Soheil Feizi}.} \bibinfo{year}{2022}\natexlab{}.
\newblock \showarticletitle{Improved Certified Defenses against Data Poisoning with (Deterministic) Finite Aggregation}. In \bibinfo{booktitle}{\emph{International Conference on Machine Learning, {ICML} 2022, 17-23 July 2022, Baltimore, Maryland, {USA}}} \emph{(\bibinfo{series}{Proceedings of Machine Learning Research}, Vol.~\bibinfo{volume}{162})}, \bibfield{editor}{\bibinfo{person}{Kamalika Chaudhuri}, \bibinfo{person}{Stefanie Jegelka}, \bibinfo{person}{Le~Song}, \bibinfo{person}{Csaba Szepesv{\'{a}}ri}, \bibinfo{person}{Gang Niu}, {and} \bibinfo{person}{Sivan Sabato}} (Eds.). \bibinfo{publisher}{{PMLR}}, \bibinfo{pages}{22769--22783}.
\newblock
\urldef\tempurl%
\url{https://proceedings.mlr.press/v162/wang22m.html}
\showURL{%
\tempurl}


\bibitem[Xu et~al\mbox{.}(2021)]%
        {MNTD}
\bibfield{author}{\bibinfo{person}{Xiaojun Xu}, \bibinfo{person}{Qi Wang}, \bibinfo{person}{Huichen Li}, \bibinfo{person}{Nikita Borisov}, \bibinfo{person}{Carl~A. Gunter}, {and} \bibinfo{person}{Bo Li}.} \bibinfo{year}{2021}\natexlab{}.
\newblock \showarticletitle{Detecting {AI} Trojans Using Meta Neural Analysis}. In \bibinfo{booktitle}{\emph{2021 IEEE Symposium on Security and Privacy (SP)}}.
\newblock


\bibitem[Yang et~al\mbox{.}(2021)]%
        {yangFeatureExtractionNovelty2021}
\bibfield{author}{\bibinfo{person}{Kun Yang}, \bibinfo{person}{Samory Kpotufe}, {and} \bibinfo{person}{Nick Feamster}.} \bibinfo{year}{2021}\natexlab{}.
\newblock \bibinfo{title}{Feature {{Extraction}} for {{Novelty Detection}} in {{Network Traffic}}}.
\newblock
\newblock
\urldef\tempurl%
\url{https://doi.org/10.48550/arXiv.2006.16993}
\showDOI{\tempurl}
\showeprint[arxiv]{2006.16993}~[cs]


\bibitem[Yang et~al\mbox{.}(2023)]%
        {Jigsaw}
\bibfield{author}{\bibinfo{person}{L. Yang}, \bibinfo{person}{Z. Chen}, \bibinfo{person}{J. Cortellazzi}, \bibinfo{person}{F. Pendlebury}, \bibinfo{person}{K. Tu}, \bibinfo{person}{F. Pierazzi}, \bibinfo{person}{L. Cavallaro}, {and} \bibinfo{person}{G. Wang}.} \bibinfo{year}{2023}\natexlab{}.
\newblock \showarticletitle{Jigsaw Puzzle: Selective Backdoor Attack to Subvert Malware Classifiers}. In \bibinfo{booktitle}{\emph{2023 IEEE Symposium on Security and Privacy (SP)}}. \bibinfo{publisher}{IEEE Computer Society}, \bibinfo{address}{Los Alamitos, CA, USA}, \bibinfo{pages}{719--736}.
\newblock
\urldef\tempurl%
\url{https://doi.org/10.1109/SP46215.2023.10179347}
\showDOI{\tempurl}


\bibitem[Zhu et~al\mbox{.}(2023)]%
        {zhuSelectiveAmnesiaEfficient2023}
\bibfield{author}{\bibinfo{person}{Rui Zhu}, \bibinfo{person}{Di Tang}, \bibinfo{person}{Siyuan Tang}, \bibinfo{person}{XiaoFeng Wang}, {and} \bibinfo{person}{Haixu Tang}.} \bibinfo{year}{2023}\natexlab{}.
\newblock \showarticletitle{Selective {{Amnesia}}: {{On Efficient}}, {{High-Fidelity}} and {{Blind Suppression}} of {{Backdoor Effects}} in {{Trojaned Machine Learning Models}}}. In \bibinfo{booktitle}{\emph{2023 {{IEEE Symposium}} on {{Security}} and {{Privacy}} ({{SP}})}}. \bibinfo{publisher}{IEEE Computer Society}, \bibinfo{pages}{682--700}.
\newblock
\showISBNx{978-1-66549-336-9}
\urldef\tempurl%
\url{https://doi.org/10.1109/SP46215.2023.00070}
\showDOI{\tempurl}


\end{thebibliography}

\newpage
\clearpage
\appendix
\section{Appendix} 

\subsection{Data features}
\label{sec:features}

For network data (CTU-13 dataset), the Zeek logs are partitioned in 30s windows and aggregated by internal IP and destination port. Each data point is comprised of several statistical features, which are summarized in~\Cref{tab:network_features}. 
The final feature count for this data representation is 1152.

\begin{table}
\centering
\caption{Statistical features for network data.}
\small
\begin{tabular}{l}
\hline
\textbf{Feature types} \\ \hline
Counts of connections per transport protocol\\
Counts of connections for each conn\_state \\
Sum, min, max over packets sent and received\\
Sum, min, max over payload bytes sent and received\\
Sum, min, max over duration of connection event \\
Counts of distinct external IPs \\
Count of distinct destination ports \\
\hline 
\end{tabular}
\label{tab:network_features}
\end{table}

For malware binary files (EMBER dataset), we leverage both metadata (e.g., version number), as well as statistical information (e.g., number of urls). The features used are presented in~\Cref{tab:binary_features}.
The resulting vectors contain 2351 features.

\begin{table}
\centering
\caption{Statistical features for binary files.}
\small
\begin{tabular}{l}
\hline
\textbf{Feature types} \\ \hline
Major and minor image version\\
Major and minor linker version\\
Major and minor operating system version\\
Minor subsystem version\\
Binary size and timestamp\\
MZ signature\\
Number of read and execute sections\\
Number of write sections\\
Number of unnamed sections\\
Number of zero-size sections\\
Counts of distinct file system paths\\
Counts of distinct registries\\
Counts of distinct URLs\\
\hline 
\end{tabular}
\label{tab:binary_features}
\end{table}

\subsection{Additional results for iterative scoring}
\label{app:iterative-scoring}

We include here visualizations for the iterative scoring process when applied to the stealthier, but less effective, versions of the backdoor poisoning attacks we consider.

\begin{figure*}[ht]
\centering
\begin{minipage}{0.45\textwidth}
  \centering
\includegraphics[width=0.95\linewidth]{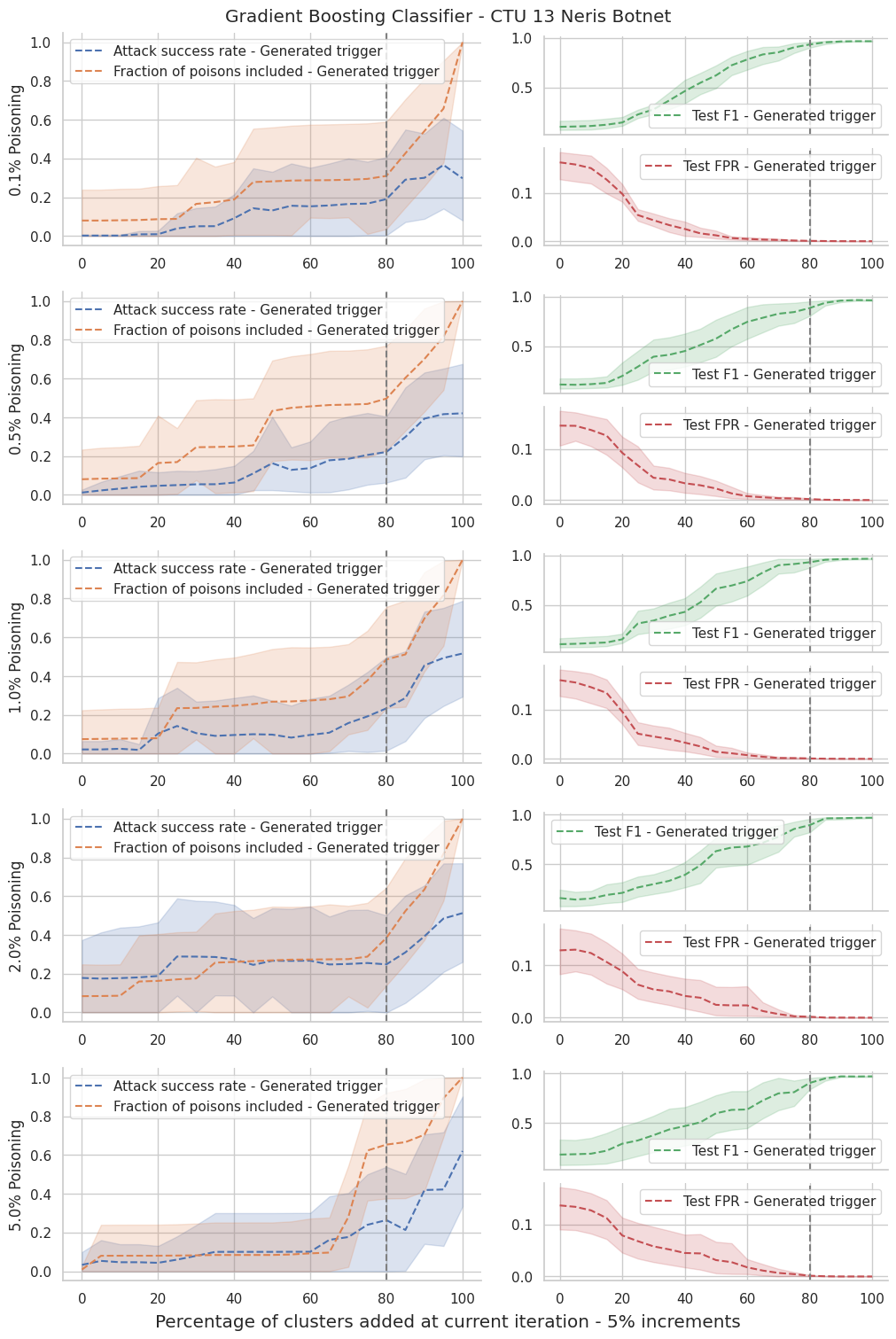}
\caption{Iterative scoring on the CTU-13 botnet classification task for the gradient boosting model. The attack was conducted with the \emph{generated trigger} strategy. The plot shows average metrics for the SHAP and Entropy attacker feature selection, at 5 different poisoning rates.}
\Description{Iterative scoring on the CTU-13 botnet classification task.}
\label{fig:gb_ctu13_iterative_gen}
\end{minipage} \quad
\begin{minipage}{0.45\textwidth}
  \centering
\includegraphics[width=0.95\linewidth]{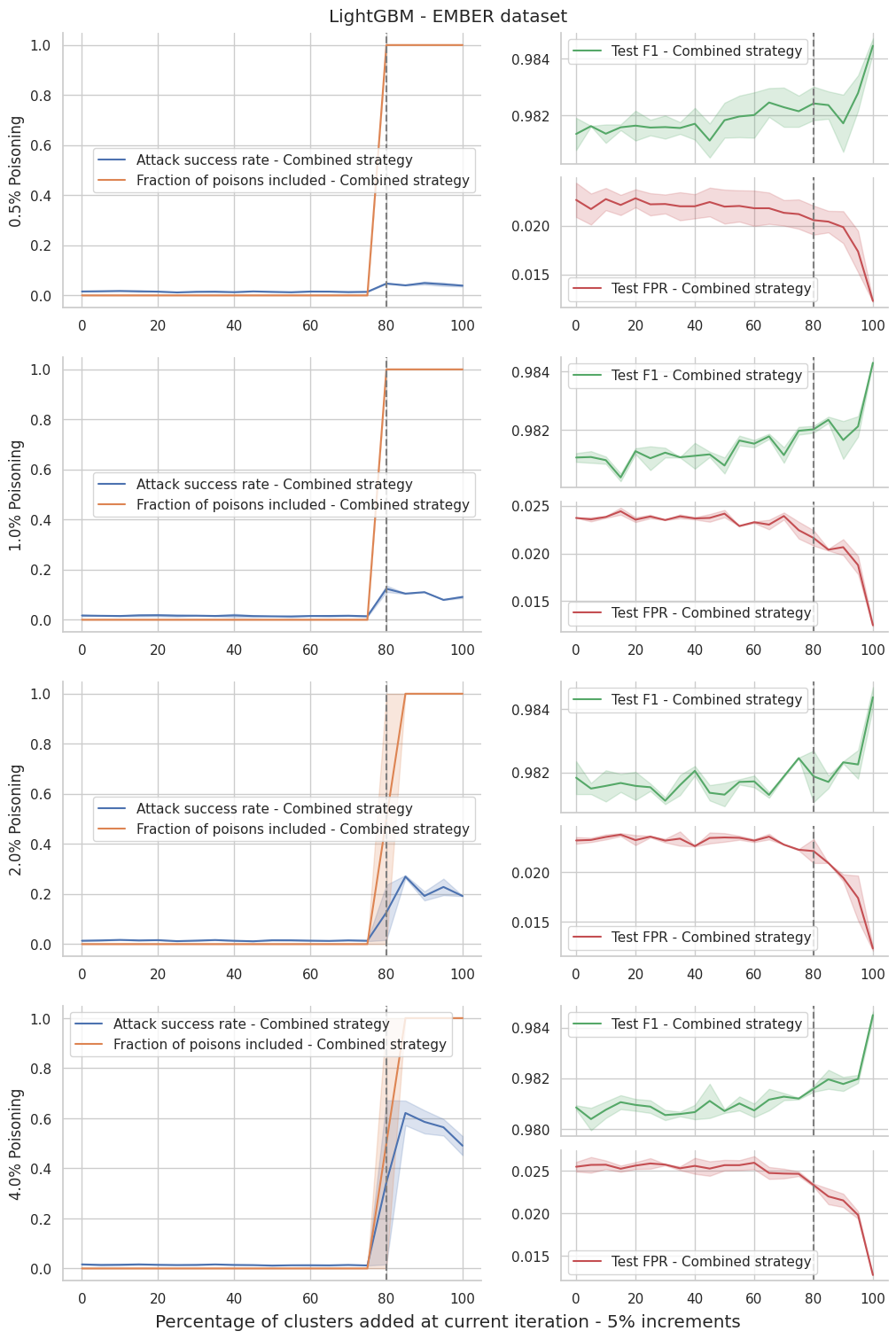}
\caption{Iterative scoring on the EMBER malware classification task for the LightGBM model. The plot shows average metrics for the set of experiments on the \emph{Combined SHAP} attack strategy at 4 different poisoning rates.}
\Description{Iterative scoring on the EMBER malware classification task.}
\label{fig:lgb_ember_iterative_combined_strat}
\end{minipage}
\end{figure*}
\pagebreak

In Figure~\ref{fig:gb_ctu13_iterative_gen} we show the dynamics of the iterative scoring process, with increments of 5\% of the clusters added to the training set, on the attack carried out with the generative trigger strategy. The vertical line indicates the fixed threshold at 80\% of included clusters.
Similarly to \Cref{fig:gb_ctu13_iterative}, each line shows the average value over runs with 5 different random seeds, and two attacker feature selection strategies: SHAP and entropy.

Figure~\ref{fig:lgb_ember_iterative_combined_strat} reports the effect of applying our iterative scoring strategy to the attack on EMBER malware classification carried out using the stealthier \emph{Combined SHAP} strategy. Each iterative step adds 5\% of the clusters, and the vertical line indicates the fixed threshold at 80\% of included clusters.
Here too, each line is an average of two observations obtained with different random seeds.

\subsection{Additional results for loss analysis and sanitization}
\label{app:loss_analysis}

We report in \Cref{tab:patching-loss-280-ctu} the effects of patching and filtering sanitization approaches when applied on the clusters identified through the analysis of the loss deltas.
The experiments are run for the gradient boosting model, with two types of trigger refinement strategies, full trigger and generated trigger.
We report the averages of 5 runs with different seeds and two feature selection strategies (SHAP, entropy), for a total of 10 experiments per value.

\begin{table}
\small
\caption{Comparison of patching and filtering sanitization approaches after loss analysis, using $z_t = 2\sigma$. Gradient boosting model on the CTU-13 dataset. Results are averages of 5 runs on different random seeds, with two feature selection approaches, for different trigger refinement strategies.}
\label{tab:patching-loss-280-ctu}

\begin{adjustbox}{width=0.98\linewidth}

\begin{tabular}{lll | c | rrr}

{\bf Sanitization} & \multicolumn{2}{c}{{\bf Attack - Poisoning}} & {\bf Base ASR} & {\bf ASR} & {\bf Test FPR} & {\bf Test F1} \\ \hline

\multirow[c]{10}{*}{Filtering} & \multirow[c]{5}{*}{Full} & 0.1\% & 50.70\% &  0.00\% &         0.01\% &       96.94\% \\
         &           & 0.5\% & 85.80\% & 30.15\% &         0.00\% &       97.07\% \\
         &           & 1\% & 88.45\% & 45.80\% &         0.00\% &       96.54\% \\
         &           & 2\% & 85.75\% & 53.00\% &         0.00\% &       96.85\% \\
         &           & 5\% & 79.90\% & 43.15\% &         0.00\% &       96.75\% \\
\cline{2-7}
         & \multirow[c]{5}{*}{Generated} & 0.1\% & 29.75\% & 35.80\% &         0.01\% &       96.80\% \\
         &           & 0.5\%  & 42.05\% & 40.95\% &         0.01\% &       96.62\% \\
         &           & 1\% & 51.65\% & 39.40\% &         0.01\% &       96.97\% \\
         &           & 2\% & 51.25\% & 46.20\% &         0.01\% &       96.82\% \\
         &           & 5\% & 62.05\% & 42.95\% &         0.01\% &       96.46\% \\

\cline{1-7}
\multirow[c]{10}{*}{Patching} & \multirow[c]{5}{*}{Full} & 0.1\% & 50.70\%   &  3.05\% &         0.00\% &       95.46\% \\
         &           & 0.5\%  & 85.80\%  & 38.05\% &         0.00\% &       96.09\% \\
         &           & 1\% & 88.45\%  & 44.20\% &         0.00\% &       95.52\% \\
         &           & 2\% & 85.75\%  & 51.70\% &         0.00\% &       96.60\% \\
         &           & 5\% & 79.90\%  & 33.85\% &         0.00\% &       95.38\% \\
\cline{2-7}
         & \multirow[c]{5}{*}{Generated} & 0.1\% & 29.75\%   & 30.25\% &         0.00\% &       96.63\% \\
         &           & 0.5\%  & 42.05\%  & 44.10\% &         0.00\% &       96.32\% \\
         &           & 1\% & 51.65\%  & 48.50\% &         0.00\% &       96.91\% \\
         &           & 2\%  & 51.25\% & 49.50\% &         0.00\% &       96.85\% \\
         &           & 5\% & 62.05\%  & 60.65\% &         0.00\% &       96.51\% \\

\end{tabular}

\end{adjustbox}

\end{table}

\subsection{Additional metrics for threshold filtering}
\label{app:numbers_at_t}

In \Cref{tab:numbers_at_t} we report a complete breakdown of the tracked utility metrics for the defended models at different fixed thresholds.
All results reported in the table are average percentages over 10 experiments, with two attacker feature selection strategies (entropy and SHAP) and 5 random seeds.

\begin{table*}

\begin{adjustbox}{width=0.85\linewidth}

\begin{tabular}{lll | rrr}
\hline
{\bf Trigger type} & {\bf Model type} & {\bf Poisoning} \% & {\bf F1 at 80.0\%} &  {\bf FPR at 80.0\% } &  {\bf Poisons in $D_{clean}$ at 80.0\%} \\
\hline

\multirow[c]{10}{*}{Full} & \multirow[c]{5}{*}{Neural Network} & 0.1\%  &        48.19\% &          1.45\% &             10.99\% \\
          &                  & 0.5\%  &        45.97\% &          1.62\% &             31.71\% \\
          &                  & 1\% &        52.31\% &          1.30\% &             13.89\% \\
          &                  & 2\% &        49.33\% &          1.36\% &             12.03\% \\
          &                  & 5\% &        64.55\% &          0.83\% &             20.51\% \\
        \cline{2-6} 
          & \multirow[c]{5}{*}{Gradient Boosting} & 0.1\%  &        85.93\% &          0.21\% &              0.00\% \\
          &                  & 0.5\%  &        94.84\% &          0.04\% &              2.72\% \\
          &                  & 1\% &        95.43\% &          0.04\% &              0.62\% \\
          &                  & 2\% &        87.13\% &          0.18\% &              0.01\% \\
          &                  & 5\% &        96.50\% &          0.02\% &              0.03\% \\
\hline
   \multirow[c]{5}{*}{Generated}       & \multirow[c]{5}{*}{Gradient Boosting} & 0.1\%  &        93.26\% &          0.07\% &             30.99\% \\
          &                  & 0.5\%  &        88.51\% &          0.18\% &             49.64\% \\
          &                  & 1\% &        93.16\% &          0.08\% &             48.59\% \\
          &                  & 2\% &        89.12\% &          0.17\% &             38.49\% \\
          &                  & 5\% &        90.10\% &          0.14\% &             65.36\% \\
\bottomrule
\end{tabular}

\end{adjustbox}

\caption{Average model utility metrics on CTU-13. Results reported for different victim architectures, at different poisoning percentages. All results are averages of 10 runs, with two attack strategies and 5 random seeds.}
\Description{Average model utility metrics on CTU-13 Neris detection expressed in percentage. Results reported for different victim architectures, at different poisoning percentages. All results are averages of 10 runs, with two attack strategies and 5 random seeds.}
\label{tab:numbers_at_t}

\end{table*}

\end{document}